\documentclass[conference]{IEEEtran}
\ifCLASSINFOpdf
\else
\fi

\usepackage{graphicx}          

\usepackage{amsfonts}       

\usepackage{amsmath,bm, bbm}
\usepackage{xcolor}

\DeclareMathOperator*{\argmin}{\mathtt{argmin}}

\newtheorem{theorem}{Theorem}
\newtheorem{corollary}{Corollary}[theorem]
\newtheorem{lemma}[theorem]{Lemma}
\newtheorem{definition}{Definition}[theorem]
\newtheorem{proposition}{Proposition}[theorem]

\DeclareMathOperator\supp{\mathtt{supp}}
\DeclareMathOperator\marg{\mathtt{marg}}

\definecolor{ao(english)}{rgb}{0.0, 0.5, 0.0}
\definecolor{americanrose}{rgb}{1.0, 0.01, 0.24}
\definecolor{cerisepink}{rgb}{0.93, 0.23, 0.51}
\definecolor{darkorchid}{rgb}{0.6, 0.2, 0.8}
\definecolor{applegreen}{rgb}{0.55, 0.71, 0.0}
\definecolor{brightpink}{rgb}{1.0, 0.0, 0.5}
\definecolor{azure(colorwheel)}{rgb}{0.0, 0.5, 1.0}
\definecolor{blue-violet}{rgb}{0.54, 0.17, 0.89}
\definecolor{amethyst}{rgb}{0.6, 0.4, 0.8}

\usepackage{datetime}

\newcommand*{\thisdraft}{This draft: April 2021} 
\newcommand*{\firstdraft}{First draft: April 2021}  
\begin{document}
%
\title{Informational Design of Dynamic Multi-Agent System}

\date{\thisdraft \\ \firstdraft}

\author{\IEEEauthorblockN{Tao Zhang and Quanyan Zhu}
\IEEEauthorblockA{Electrical and Computer Engineering\\
New York University\\
Email: \{tz636, qz494\}@nyu.edu}\\
\IEEEauthorblockN{\normalsize{First Draft: April, 2021}\\
\normalsize{This Draft: June 2021}}
}


%


\maketitle

\begin{abstract}
This work considers a novel information design problem and studies how the craft of payoff-relevant environmental signals solely can influence the behaviors of intelligent agents.
The agents' strategic interactions are captured by a Markov game, in which each agent first selects one external signal from multiple signal sources as additional payoff-relevant information and then takes an action.
There is a rational information designer (principal) who possesses one signal source and aims to influence the equilibrium behaviors of the agents by designing the information structure of her signals sent to the agents.
We propose a direct information design approach that incentivizes each agent to select the signal sent by the principal, such that the design process avoids the predictions of the agents' strategic selection behaviors. 
We then introduce the design protocol given a goal of the designer which we refer to as obedient implementability (OIL) and characterize the OIL in a class of obedient sequential Markov perfect equilibria (O-SMPE).
A design regime is proposed based on an approach which we refer to as the fixed-point alignment that incentivizes the agents to choose the signal sent by the principal, guarantees that the agents' policy profile of taking actions is the policy component of an O-SMPE and the principal's goal is achieved.
We then formulate the principal's optimal goal selection problem in terms of information design and characterize the optimization problem by minimizing the fixed-point misalignments.
The proposed approach can be applied to elicit desired behaviors of multi-agent systems in competing as well as cooperating settings and be extended to heterogeneous stochastic games in the complete- and the incomplete-information environments.
\end{abstract}


%
\IEEEpeerreviewmaketitle

\section{Introduction}

Building rational multi-agent system is an important research desideratum in Artificial Intelligence.
In goal-directed decision making systems, an agent's action is controlled by its consequence \cite{dickinson1985actions}.
In a game, the consequence of an agent's action is the outcome of the game, given as the reward of taking that action as well as the actions of his opponents, which situates the optimality criterion of each agent's decision making in the game.
A rational agent's reward may also depend on the payoff-relevant information, in addition to the actions.
The information may include the situation of the agents in a game, referred to as the state of the world, as well as his knowledge about his opponents' diverging interests and their preferences over the outcomes of the game.
Incorporating such payoff-relevant information in his decisions constitutes an essential part of an agent's rationality in the strategic interactions with his opponents.
Hence, one may re-direct the goal achievement of rational agents in a game by information provision.
In economics, this refers to as \textit{information design}, which studies how an information designer (she) can influence agents' optimal behaviors in a game to achieve her own objective, through the design of information provided to the game \cite{bergemann2019information}.

Referred to as the inverse game theory, mechanism design is a well-developed mathematical theory in economics that provides general principles of how to 
design \textit{rules of games} (e.g., rewarding systems with specifications of actions and outcomes) to influence the agents' strategic interactions and achieve system-wide goals while treating the information as given.
Information design, on the other hand, considers the circumstances when the information in the environment is under the control of the system designer and offers a new approach to indirectly elicit agents' behaviors by keeping the game rules fixed \cite{taneva2019information}.
%

This work considers a infinite-horizon Markov game of a finite number of agents.
Each agent's identity is characterized by his type.
%
%
At each period of time, agents observe a payoff-relevant global state (state).
In addition to the state, each agent observes a batch of signals (signal batch, batch) at each period and then strategically chooses one signal from the batch as the additional information to support his decision of  taking a action.
Each agent's one-period reward (parameterized by the type) is determined by his own action, the actions of his opponents, the global state, and his choice of signal.
We refer to this game as a \textit{base Markov game} (BMG).
The transition of the state and the distribution of signals are referred to as the \textit{information structure} of the BMG.
%
%
In a BMG, each agent's behavior includes selecting a signal according to a \textit{selection rule} and taking an action according to a \textit{policy}.
Here, each agents' selection of signal and the choice of action are coupled since the selected signal enters the policy to determine the choice of the action.
If a mechanism designer aims to incentivize the agents to behave in her desired way, she directly modifies the BMG--\textit{reversing the game}--by designing the game model, including changing the reward function associated with actions and outcomes, while treating the information structure as a given part of the environment.
%
%
An information designer, however, treats the BMG model as fixed and modifies the information structure to elicit agents' equilibrium behaviors that coincide with her objective.

We study a novel dynamic information design problem in the BMG in which there are multiple sources of signals (signal sources, sources) and each of them sends one signal to each agent.
The signals sent by all sources constitute the signal batch observed by each agent at each time.
Among these sources, there is one rational information designer (referred to as \textit{principal}, she) who controls one signal source and intends to strategically craft the information structure of her signal by choosing a \textit{signaling rule} to indirectly control the equilibrium of the BMG.
We consider that other sources of signals provide additional information to the agents in a non-strategic take-it-or-leave-it manner.
%
%
The goal of the principal is to induce the agents to take actions according to an equilibrium policy that is desired by the principal.
However, the principal has no ability to directly program the agents' behaviors to force them to take certain actions.
Instead, her information design should provide incentive to rational agents to behave in her favor.
We study the extent to which the provision of signals along by controlling a single signal source can influence the agents' behavior in a BMG, when the agents have the freedom to choose any available signal in the batch.
We will name the BMG with a rational principal in this setting as an \textit{augmented Markov game} (AMG). 

Since the principal's design problem keeps the base game unchanged, our model fits the scenarios when the agents are intrinsically motivated and their internal reward systems translate information from external environment into internal reward signals \cite{chentanez2005intrinsically}. 
Intrinsically-motivated rational agents can be human decision makers with intrinsic psychological preferences or intelligent agents programmed with internal reward system.
The setting of multiple sources of additional information captures the circumstances when the environment is perturbed by noisy information, in which the agents may improperly use redundant and useless information to make their decisions that may deviate from the system designer's desire.
Also, the principal can be an adversary who aims to manipulate the strategic interactions in a multi-agent system through the provision of disinformation, without intruding each agent's local system to make any physical or digital modifications.

Although the principal's objective of the information design in an AMG is to elicit an equilibrium distribution of actions, her design problem has to take into consideration how the agents select the signals from their signal batches because each agent's choice of action is coupled with his selection of signal. 
In an information design problem, the principal chooses an information structure such that each agent selects a signal using a selection rule and then takes an action according to a policy which matches the principal's goal.
The latter is constrained by the notion \textit{admissibility}.
%
%
In general, the signals sent by the principal may not be selected by some agents, thereby the actions taken by those agents are independent of the principal's (realized) signals.
However, even though her signal does not enter an agent's policy to realize an action, the principal may still influence the agent's action because the information structure of the signal batch is influenced by her choice of signaling rule.
The information structure of the signal batch affects the agents' selections of signals.
Hence, the agents' behaviors indirectly depend on the principal's signaling rule.
We refer to such information design as \textit{indirect information design} (IID) which requires the principal to accurately predict each agents' strategic selection rule and their policy profiles that might be induced by the signaling rule.
We restrict attention to another class of information design, referred to as \textit{direct information design} (DID).
In DID problems, each agent always selects the signal sent by the principal and then takes an action.
Thus, the realizations of the principal's signals directly enter the agents' policies to choose actions.
In addition to the admissibility, another restriction of the principal's DID problem is captured by the notion of \textit{obedience} which requires that each agent is incentivized to select the signal from the principal rather than choose one from other signal sources.
The key simplification provided by the DID is that the principal's prediction of the agents' strategic selection rules is replaced by replaced by a straightforward obedient selection rule that always prefers the principal's signals.

This paper makes three major contributions to the foundations of information design.
%
First, we define a dynamic direct information design problem in an environment where the agents have the freedom to choose any available signal as addition payoff-relevant information.
Captured by the notion of \textit{obedient implementability}, the principal's information problem is constrained by the \textit{obedient} condition that incentivizes the agents to prefer the signals sent by the principal and the \textit{admissibility} condition such that the agents take actions which meets the principal's goal.
Our information design problem is distinguished from others in economics that study the commitment of the information design in a game when there is only a single source of additional information in static settings (e.g., \cite{mathevet2020information,taneva2019information,bergemann2016bayes,kamenica2011bayesian}) as well as in dynamic environment (e.g., \cite{ely2015suspense,passadore2015robust,doval2020sequential,ely2017beeps,ely2020moving,makris2018information}) and the settings in which the agents do not make a choice from multiple designers (e.g., \cite{koessler2018interactive}).

Second, we propose a new solution concept termed \textit{obedient sequential Markov perfect equilibrium} (O-SMPE) which allows us to handle the undesirable deviations of agents in a principled manner.
By bridging the augmented Markov game model with dynamic programming and uncovering the close relationship between the sequential-perfect relationship of the O-SMPE and a pair of fixed points, we characterize the obedient implementability and explicitly formulate the information design regime given a goal of the principal.
The proposed framework is based on an approach referred to as fixed-point alignment which selects a signaling rule that matches the first fixed point from the agents' optimal signaling selection to the second fixed point of optimal action takings.
However, the principal cannot achieve just any goal she wishes to.
We identify the key conditions, known as Markov perfect goal and strong admissibility, that discipline the freedom of the principal's ability to influence the agents' equilibrium behaviors.

Third, we formulate the principal's goal selection problem and transform it to a direct information design problem without a predetermined goal.
The principal's problem is thus to select a signaling rule such that it induces agents' equilibrium policy profiles that maximize (optimal information design) or minimize (robust information design) her expected payoff.
Our formulation does not assume the availability of all possible equilibrium policy profiles that can be induced by each possible signaling rule and thus the principal can select equilibrium policy profiles for her choice of signaling rule.
Instead, our framework takes into consideration the role of the signaling rule in ensuring the agents to converge to an equilibrium.
A new approach is proposed based on a condition known as the fixed-point misalignments minimization that captures a new optimality criterion for the information design without a given goal.

\subsection{Related Work}

We follow a growing line of research on creating incentives for interacting agents to behave in a desired way.
The most straightforward way is based on mechanism design approaches that properly provide reward incentives (e.g., contingent payments, penalty, supply of resources) by directly modifying the game itself to change the induced preferences of the agents over actions.
Mechanism design approaches have been fruitfully studied in both static \cite{myerson1981optimal} as well as dynamic environment \cite{pavan2014dynamic,zhang2021incentive,zhang2021differential}.
For example, auctions \cite{milgrom2004putting,bhat2019optimal} specify the way in which the agents can place their bid and clarify how the agents pay for the items; in matching markets \cite{sonmez2011matching,zhang2019optimal}, matching rules matches agents in one side of a market to agents of another side that directly affect the payoff of each matched individuals.
%
%
In reinforcement learning literature, reward engineering \cite{dewey2014reinforcement,nagpal2020reward,hadfield2017inverse} is similar to mechanism design that directly crafts the reward functions of the agents that post specifications of the learning goal.

Our work lies in another direction: the information design. 
Information design studies how to influence the outcomes of the decision makings by choosing signal (also referred to as signal structure, information structure, Blackwell experiment, or data-generating process) whose realizations are observed by the agents \cite{kamenica2019bayesian}.
In a seminal paper \cite{kamenica2011bayesian}, Kamenica and Gentzkow has introduced \textit{Bayesian persuasion} in which there is an informed sender and an uninformed receiver. 
%
The sender is endowed to commit to choosing any probability distribution (i.e., the information structure) of the signals as a function of the state of the world which is payoff-relevant to and unobserved by the receiver.
The Bayesian persuasion can be interpreted as a communication device that is used by the sender to inform the receiver through the signals that contain knowledge about the state of the world. Hence, the sender controls what the agent gets to know about the payoff-relevant state.
With the knowledge about the information structure, the receiver forms a posterior belief about the unobserved state based on the received signal. 
Hence, the information design of Bayesian persuasion is also referred to as an exercise in belief manipulation.
Other works alongside with the Bayesian persuasion include \cite{brocas2007influence,rayo2010optimal,arieli2019private,castiglioni2020online}.
In \cite{mathevet2020information}, Mathevet et al. extends the single-agent Bayesian persuasion of \cite{kamenica2011bayesian} to a multi-agent game and formulate the information design of influencing agents' behaviors through inducing distributions over agents' beliefs.
In \cite{bergemann2016bayes}, Bergemann and Morris have also considered information design in games. They have formulated the Myersonian approach for the information design in an incomplete-information environment. 
The essential of the Myersonian information design is the notion of Bayes correlated equilibrium, which characterizes the all possible Bayesian Nash equilibrium outcomes that could be induced by all available information structures.
The Myersonian approach avoids the modeling of belief hierarchies \cite{mertens1985formulation} and constructs the information design problem as a linear programming.
Information design has been applied in a variety of areas to study and improve real-world decision making protocols, including stress test in finance \cite{goldstein2018stress,inostroza2018persuasion}, law enforcement and security  \cite{hernandez2018bayesian,rabinovich2015information}, censorship \cite{gehlbach2014government}, routing system \cite{das2017reducing}, finance and insurance \cite{duffie2017benchmarks,szydlowski2021optimal,garcia2021information}.
Kamenica \cite{kamenica2019bayesian} has provided a recent survey of the literature of Bayesian persuasion and information design.

This work fundamentally differs from existing works on the information design.
First, we consider a different environment.
Specifically, we consider the setting when there are multiple sources of signals and each agent chooses one realized signal as an additional (payoff-relevant) information at each time.
Among these sources of signals, there is an information designer who controls one of these sources and aims to induce equilibrium outcomes of the incomplete-information Markov game by strategically crafting information structures.
Second, other than only taking actions, each agent in our model makes a coupled decision of selecting a realized signal and taking an action.
Hence, the characterization of the solution concepts in our work is different from the equilibrium analysis in other works.
Third, we also provide an approach with an explicit formulation to relaxing the optimal information design problem.

In this section, we first describe some fundamental concepts of a canonical Markov game model and then define our new model of a game called \textit{augmented Markov game} by extending the canonical model.

\noindent \textbf{Conventions.} For any measurable set $X$, $\Delta(X)$ is the set of probability measures over $X$. Any function defined on a measurable set is assumed to be measurable.
For any distribution $P\in \Delta(X_{1} \times X_{2})$, for two measurable sets $X_{1}$ and $X_{2}$, $\marg_{X_{1}} P$ is the marginal distribution of $P$ over $X_{1}$ and $\supp P$ is the support of $P$.
The history of $x\in X$ from period $s$ to period $t$ is denoted by $x^{(t)}_{s}$ with $x^{(t)}$ when $s=0$.
We use $P_{r}(E)$ to represent the probability of an event $E$.
For the compactness of notations, we only show the elements, but not the sets, over which are summed under the summation operator. 
The notations are summarized in Appendix \ref{app:list_of_notations}.

\subsection{Normal-Form Game and Canonical Markov Game}

This work considers games of finite agents, denoted by $\mathcal{N}\equiv[n]$, $0<n<\infty$.
%
%
A \textit{normal-form} (or strategic-form) is a basic representation of a static game:
\begin{definition}[Normal-Form Game \cite{ziebart2011maximum}]\label{def:normal_form_game}
A normal-form game is defined by a tuple $G\equiv<\mathcal{N}, \mathcal{A},  \{r_{i}\}_{i\in\mathcal{N}}>$.
%
Here, $\mathcal{A}$ is a finite set of actions available to each agent. 
$r_{i}:\mathcal{A}^{n}\mapsto \mathbb{R}$ is the reward function of agent $i$.
\end{definition}

%
Each agent $i\in \mathcal{N}$ simultaneously chooses an action $a_{i}\in \mathcal{A}$ and receives a reward $r_{i}(a_{i}, \bm{a}_{-i})$ when other agents choose actions $\bm{a}_{-i}\in\mathcal{A}^{n-1}$.

\begin{figure}[h]\label{fig:canonical_Markov_game}
\includegraphics[width=\linewidth]{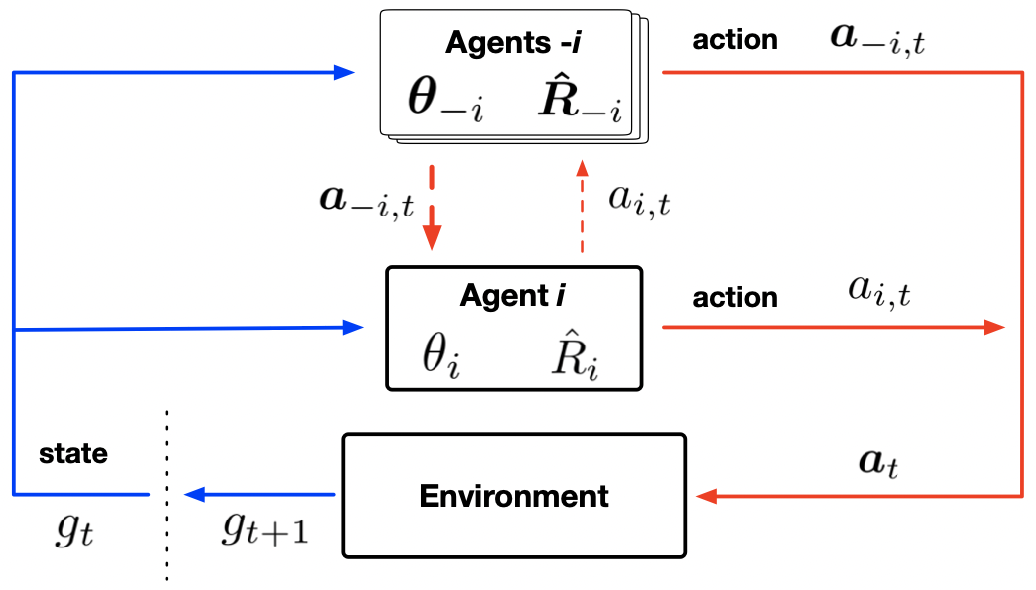}
\caption{Canonical Markov game.}
\centering
\end{figure}

With reference to Fig. \ref{fig:canonical_Markov_game}, a canonical Markov game generalizes a normal-form game to dynamic settings as well as Markov decision processing to multi-agent interactions.
We consider a finite-agent infinite-horizon Markov game. The game is played in discrete time indexed by $t=0,1,\dots$.
Each agent $i$'s identification is captured by the parameter known as \textit{type} denoted by $\theta_{i}$.
A canonical Markov game can be defined by a tuple $\widehat{M}[\bm{\theta}]\equiv<\mathcal{N}, \mathcal{G},\mathcal{A}, d_{g}, \mathcal{T}_{g}, \{\hat{R}_{i}(\cdot|\theta_{i})\}_{i\in\mathcal{N}}>$.
%
%
Here, $\mathcal{G}$ is a finite set of states. $\mathcal{A}$ is a finite set of actions available to each agent at each period.
$d_{g}\in\Delta(\mathcal{G})$ is an initial distribution of the state. $\mathcal{T}_{g}: \mathcal{G}\times \mathcal{A}^{n} \mapsto \Delta(\mathcal{G})$ is the transition function of the state, such that $\mathcal{T}_{g}(\cdot|g_{t}, \bm{a}_{t})\in \Delta(\mathcal{G})$ specifies the probability distribution of next-period state when the current state is $g_{t}$ and the current-period joint action is $\bm{a}_{t}\equiv(a_{i,t})_{i\in\mathcal{N}}$.
Each agent $i$'s goal is characterized by his reward function $\hat{R}_{i}(\cdot|\theta_{i}): \mathcal{G}\times \mathcal{A}^{n}\mapsto \mathbb{R}$ is the reward function of agent $i$ parameterized by $\theta_{i}\in\Theta$ that realizes a one-stage reward $r_{i}(g_{t}, \bm{a}_{t})$ for agent $i$ when the state is $g_{t}$ and the joint action is $\bm{a}_{t}$.
A canonical Markov game is complete-information.

A solution to $\widehat{M}$ is a sequence of \textit{policy profile} $\{\bm{\hat{\pi}}_{t}\}_{t\geq 0}$ in which $\bm{\hat{\pi}}_{t}: \mathcal{G} \times \mathcal{A}^{n} \mapsto \Delta(\mathcal{A}^{n})$, such that $\bm{\hat{\pi}}_{t}(\cdot| g_{t})$ specifies the probability of the joint action of the agents given the current-period state $g_{t}$.
A policy profile is \textit{stationary} if the agents' decisions of actions depend only on the current-period payoff-relevant information and is independent of the calendar time; i.e., we denote the policy profile as $\bm{\hat{\pi}}: \mathcal{G} \times \mathcal{A}^{n} \mapsto \Delta(\mathcal{A}^{n})$.
When the policy profile is a \textit{pure} strategy if $\bm{\hat{\pi}}: \mathcal{G} \times \mathcal{A}^{n} \mapsto \mathcal{A}^{n}$ is a deterministic mapping. 
In a canonical Markov game, $\bm{\hat{\pi}}$ can be either \textit{independent} (i.e., $\bm{\hat{\pi}}(\bm{a}_{t}| s_{t})=\prod_{i\in \mathcal{N}} \hat{\pi}_{i}(a_{i,t}|s_{t})$) or \textit{correlated} (i.e., a joint function).

%
According to Ionescu Tulcea theorem (see, e.g., \cite{hernandez2012discrete}), the initial distribution $d_{g}$ on $g_{0}$, the transition function $\mathcal{T}_{g}$, and the policy profile $\bm{\hat{\pi}}$ together define a unique probability measure $P^{\bm{\hat{\pi}}}$ on $(\mathcal{G} \times \mathcal{A}^{n})^{\infty}$. We denote the expectation with respect to $P^{\bm{\hat{\pi}}}$ as $\mathbb{E}_{\bm{\hat{\pi}}}[\cdot]$.
The optimality criterion for each agent's decision making at each period is to maximize his expected payoff. Each agent discounts his future payoffs by a discount factor $0<\gamma<1$. 
Thus, agent $i$'s period-$t$ infinite-horizon \textit{interim expected payoff} is defined as: for $g_{t}\in\mathcal{G}$, $\bm{a}_{t}\in\mathcal{A}^{n}$, $\theta_{i}\in\Theta$, $i\in\mathcal{N}$, $t\geq 0$, 
\begin{equation}\label{eq:agent_expected_payoff_preliminary}
    \begin{aligned}
    \mathtt{Expr}_{i}(g_{t},\bm{a}_{t};\theta_{i} |\bm{\hat{\pi}}, \hat{R}_{i})\equiv\mathbb{E}_{\bm{\hat{\pi}}}\Big[\sum_{s=t}^{\infty}\gamma^{t} \hat{R}_{i}(\bm{a}_{s},g_{s}|\theta_{i})\Big| g_{t}, \bm{a}_{t} \Big].
    \end{aligned}
\end{equation}

\begin{definition}[MPE]
A policy profile $\bm{\hat{\pi}}^{*}$ constitutes a \textbf{stationary Markov perfect equilibrium} (MPE) if, $\bm{\hat{\pi}}^{*}$ is independent, and for $a_{i,t}\in\mathcal{A}$ with $\hat{\pi}^{*}_{i}(a_{i,t}|g,\theta_{i})>0$, $a'_{i,t}\in\mathcal{A}$, $g\in\mathcal{G}$, $t\geq 0$, $i\in\mathcal{N}$,
\begin{equation}\label{eq:canonical_MPE_condition}
    \begin{aligned}
    \mathbb{E}_{\bm{\hat{\pi}}^{*}_{-i}}\Big[\mathtt{Expr}_{i}&(g_{t}, a_{i,t}, \bm{a}_{-i,t};\theta_{i} |\bm{\hat{\pi}}, \hat{R}_{i}) \Big] \\
    &\geq \mathbb{E}_{\bm{\hat{\pi}}^{*}_{-i}}\Big[\mathtt{Expr}_{i}(g_{t}, a'_{i,t}, \bm{a}_{-i,t};\theta_{i} |\bm{\hat{\pi}}, \hat{R}_{i}\Big].
    \end{aligned}
\end{equation}
\end{definition}

%

Let $\bm{\hat{h}}\equiv (\bm{a}^{(\infty)}, g^{(\infty)})\in \bm{\hat{H}}\equiv \mathcal{A}^{n\times \infty} \times \mathcal{G}^{\infty}$ denote any infinite sequence of action-state pairs.
Define $\mathtt{ExR}_{i}(\bm{\hat{h}})\equiv \sum_{t=0}^{\infty} \gamma^{t} \hat{R}_{i}(\bm{a}_{t}, g_{t}|\theta_{i})$, for any $\bm{\hat{h}}\in \bm{\hat{H}}$, $\theta_{i}$, $i\in\mathcal{N}$.
Let $\bm{\hat{h}}(t)$ denote the first $t$ components of $\bm{\hat{h}}$, for all $t\geq 0$.

\begin{lemma}[\cite{maskin2001markov})]\label{lemma:continuous_infinity_canonical_markov}
For any $0<\gamma<1$, The game $\widehat{M}$ is continuous at infinity; i.e., %
\begin{equation*}
    \begin{aligned}
    \lim\limits_{T\rightarrow \infty}\sup\limits_{\substack{i, \bm{\hat{h}}, \bm{\hat{h}'},  \\ \bm{\hat{h}}(T-1) =  \bm{\hat{h}'}(T-1)}}\Big|  \mathtt{ExR}_{i}(\bm{\hat{h}}) - \mathtt{ExR}_{i}(\bm{\hat{h}'}) \Big| \rightarrow 0.
    \end{aligned}
\end{equation*}
\end{lemma}

Lemma \ref{lemma:continuous_infinity_canonical_markov} shows that for a fixed discount rate $0<\gamma<1$, the canonical Markov game is continuous at infinity. 
The continuity at infinity is essential for the existence of MPE of infinite-horizon stochastic game.
The following proposition shows the existence of MPE due to \cite{maskin2001markov}.

\begin{proposition}[\cite{maskin2001markov}]
Suppose that $0<\gamma<1$. Then, the game $\widehat{M}$ admits a MPE. 
\end{proposition}

\subsection{Augmented Markov Game Model}

\begin{figure*}[h]\label{fig:augmented_Markov_game}
\includegraphics[width=\linewidth]{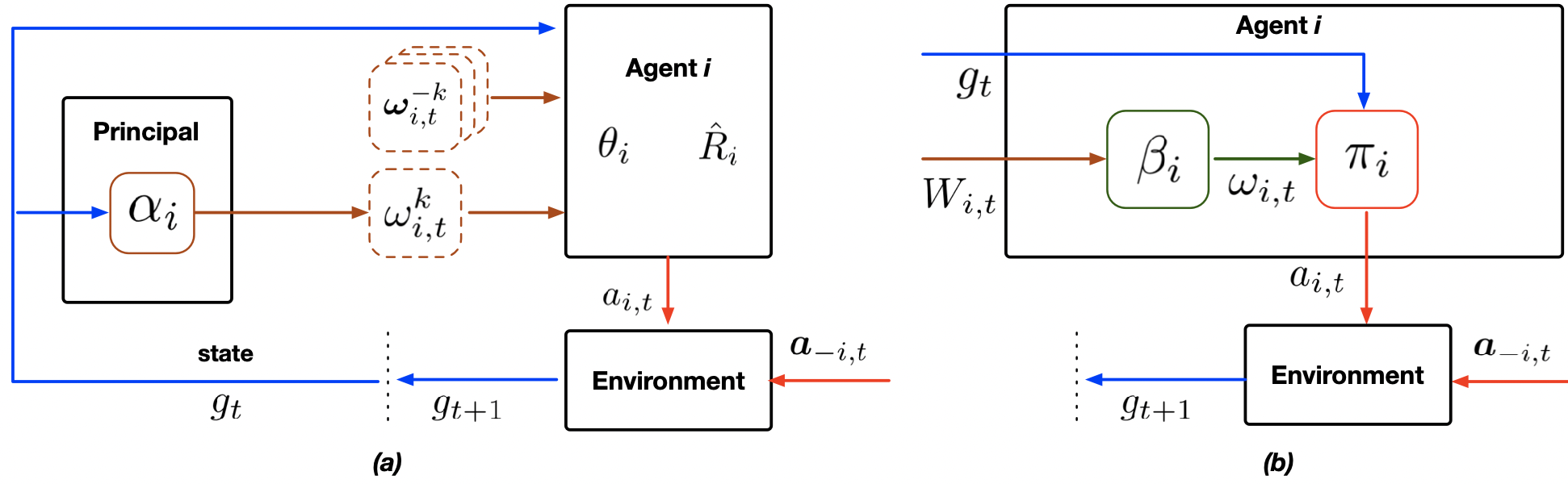}
\caption{Augmented Markov game. \textit{(a)} shows external activities of the augmented game from agent $i$'s point of view with a principal and \textit{(b)} describes agent $i$'s internal decision making processes.
Specifically, each agent $i$ received a state $g_{t}$ and a signal batch $W_{i,t}=\{\omega^{k}_{i,t},\bm{\omega}^{-k}_{i,t}\}$ in which the signal $\omega^{k}_{i,t}$ is chosen by the principal using the signaling rule $\alpha_{i}$ that depends on the state $g_{t}$.
Each agent $i$ first uses $\beta_{i}$ to select one signal $\omega_{i,t}$ from $W_{i,t}$. By taking into account the state $g_{t}$ and his selection $\omega_{i,t}$, he uses $\pi_{i}$ to take an action $a_{i,t}$. The state $g_{t}$ is transitioned to a next state $g_{t+1}$ given the agents' joint action $\bm{a}_{t}=\{a_{i,t},\bm{a}_{-i,t}\}$.   } 
\centering
\end{figure*}

With reference to Fig. \ref{fig:augmented_Markov_game}, we extend the canonical Markov game model in this section by introducing an additional payoff-relevant information for the agents in addition to the state.

\textbf{Signals.} We call the additional information as \textit{signal}.
Let $\Omega$ be a finite set of signals.
We consider that at each period $t$ each agent $i$ observes a batch of $m$ (finite) signals (signal batch, batch), for $m>1$, denoted by $W_{i,t}\equiv\{\omega^{j}_{i,t}\}_{j=1}^{m}\subseteq \Omega^{m}$, for all $i\in\mathcal{N}$, $t\geq 0$, in which $\omega^{j}_{i,t}\in \Omega$ denotes a typical signal indexed by $j\in[m]$ sent to agent $i$ at $t$. 
Upon observing $W_{i,t}$, each agent $i$ selects one signal $\omega_{i,t}$ from the batch $W_{i,t}$ and $\omega_{i,t}$ becomes payoff-relevant to him in addition to the state $g_{t}$.

\textbf{Dynamics of Information.} Let $\omega^{k}_{i,t}$ denote one signal contained in the batch $W_{i,t}$.
Given $\omega^{k}_{i,t}$, the signal batch can be represented as $W_{i,t}=\{\omega^{k}_{i,t}, W^{-k}_{i,t}\}$, where $W^{-k}_{i,t}\equiv\{\omega^{j}_{i,t}\}_{j=1, j\neq k}^{m}$, for all $i\in\mathcal{N}$, $t\geq 0$.
We assume that the probability of $\omega^{k}_{i,t}$ at any $t$ is specified by $\mathcal{P}^{k}_{i,t}\in \Delta(\Omega)$. 
Let $\mathcal{P}^{-k}_{i,t}\in \Delta(\Omega^{m-1})$ denote the probability of $W^{-k}_{i,t}$.
Here, both $\mathcal{P}^{k}_{i,t}$ and $\mathcal{P}^{-k}_{i,t}$, for all $i\in\mathcal{N}$ may depend on the current state $g_{t}$, the joint type $\bm{\theta}$, the calendar time, or the past realizations of states or actions.
Although there is an additional information in the game, the transition of the state is controlled by the current $g_{t}$ and the realized actions $\bm{a}_{t}$ and, however, is independent of the selected signal $\omega_{i,t}$, for all $i\in\mathcal{N}$, $t\geq 0$.
As in the canonical game $\widehat{M}$, the transition of the state is Markov: given any $g_{t}\in\mathcal{G}, \bm{a}_{t}\in\mathcal{A}^{n}$, the transition function $\mathcal{T}_{g}(g_{t+1}|g_{t}, \bm{a}_{t})$ specifies the probability of a next state $g_{t+1}\in \mathcal{G}$, for all $t\geq 0$.

\textbf{Augmented Markov Game.} We refer to the Markov game with signal as the \textit{augmented Markov game}, denoted by $M$:
$M\equiv<\mathcal{N}, \mathcal{G},\mathcal{A}, \Omega, d_{g},  \mathcal{T}_{g},$  $\{\mathcal{P}^{k}_{i,t},\mathcal{P}^{-k}_{i}\}_{i\in\mathcal{N}, t\geq 0},$  $ \{R_{i}(\cdot|\theta_{i})\}_{i\in\mathcal{N}}>$,
where $R_{i}(\cdot|\theta_{i}):\mathcal{A}^{n} \times\mathcal{G}\times  \Omega \mapsto \mathbb{R}$ is the reward function ($\theta_{i}$-parameterized) of agent $i$, which takes into consideration agent $i$'s selected signal.
A special setting for $M$ is that each agent $i$ makes two sequentially-coupled decisions. Specifically, agent $i$ first selects a signal $\omega_{i,t}$ from the batch $W_{i,t}$, given the state $g_{t}$; then, he takes an action $a_{i,t}$, given $g_{t}$ and his selected signal $\omega_{i,t}$.
Hence, at a given period $t$, if the state is $g_{t}$, agent $i$ observes a set of signals $W_{i,t}$, the joint action of all other agents is $\bm{a}_{-i,t}$, and if agent $i$ firstly selects a signal $\omega_{i,t}$ from $W_{i,t}$ and secondly takes an action $a_{i,t}$, then the single-period reward is $R_{i}(\{a_{i,t},\bm{a}_{-i,t}\}, g_{t}, \omega_{i,t}|\theta_{i})$.
We assume that the game model $M$ is complete-information.

\subsubsection{Strategies}

Each agent $i$ is \textit{rational} in the sense that it is self-interested and makes his decisions according to his observation $(g_{t},  W_{i,t})$ to maximize his expected payoffs.
We consider that the solution to the game $M$ is a stationary Markov strategy profile $<\bm{\beta}, \bm{\pi}>$ where $\bm{\beta}: \mathcal{G} \mapsto \bm{W}_{t}$ is a \textit{selection rule profile} and 
$\bm{\pi}(\cdot|\theta^{n}): \mathcal{G}\times \Omega^{n} \times \Theta^{n} \mapsto \Delta(\mathcal{A}^{n})$ is a \textit{policy profile}.
Similar to $\bm{\hat{\pi}}$ in $\widehat{M}$, each strategy of the profiles $\bm{\beta}$ and $\bm{\pi}$ can be either \textit{correlated} (i.e, a joint function) or \textit{independent} (i.e.,  $\omega_{i,t}=\beta_{i}(g_{t})$, for all $i\in \mathcal{N}$, and $\bm{\pi}(\bm{a}_{t}|g_{t}, \bm{\omega}_{t}, \bm{\theta}_{t}) = \prod_{i\in N} \pi_{i}(a_{i,t}|g_{t}, \omega_{i,t}, \theta_{i,t})$, where $\omega_{i,t} = \beta_{i}(g_{t}, \theta_{i,t})$).

Given any observation $(g_{t}, W_{i,t}, \theta_{i,t})$, each agent $i$'s selection of the signal and his choice of the action are fundamentally different.
Specifically, the payoff-relevant information for signal selection is $g_{t}$, i.e., $\bm{\beta}(g_{t}, \bm{\theta}_{t}) \in \bm{W}_{t}$, while the payoff-relevant information for the action taking is $(g_{t}, \bm{\omega}_{t})$, i.e., $\bm{\pi}(\bm{a}_{t}|g_{t}, \bm{\omega}_{t}, \bm{\theta}_{t})\in \Delta(\bm{\mathcal{A}})$ in which $\bm{\omega}_{t} = \bm{\beta}(g_{t}, \bm{\theta}_{t})$.
However, we will write $\omega_{i,t} = \beta_{i}(g_{t},  \theta_{i,t}|W_{i,t})\in W_{i,t}$ to highlight the influence of the signal batch $W_{i,t}$ (and thus its distribution) on each agent $i$'s decision of selecting a signal.
Agent $i$ first uses $\beta_{i}$ to select signal $\omega_{i,t} = \beta_{i}(g_{t},  \theta_{i,t}|W_{i,t})$ and then chooses an action $a_{i,t}$ according to $\pi_{i}(a_{i,t}|g_{t}, \omega_{i,t}, \theta_{i,t})$ based on the realized selection $\omega_{i,t}$.

\section{Information Design Problem}

In this work, we are interested in when there is one rational \textit{information designer} referred to as \textit{principal} (she, indexed by $k$) who controls one of $m$ signal sources.
At time $t$, the signal sent to agent $i$ by the principal is denoted by $\omega^{k}_{i,t}$.
We assume that $\bm{\mathcal{P}}^{-k}_{t}\equiv\bm{\mathcal{P}}^{-k}=\{\mathcal{P}^{-k}_{i}\}_{i\in\mathcal{N}}$ is fixed and purely exogenous; i.e., $\bm{\mathcal{P}}^{-k}$ is independent of the state, types, actions, the calendar time, or the histories of the game.
This can be interpreted as when other $m-1$ sources of signals provide additional information to the agents in a non-strategic take-it-or-leave-it manner.

We consider that the principal is \textit{rational} in that she possesses a goal specified by the agents' equilibrium actions and strategically designs the information structure of her signal to achieve her goal.
Specifically, the principal aims to induce the agents to take actions that coincides with her goal in the equilibrium by strategically chooses $\bm{\mathcal{P}}^{k}_{t}\equiv\{\mathcal{P}^{k}_{i,t}\}_{i\in\mathcal{N}}$ given $\Omega$. 
Since $\bm{\mathcal{P}}^{k}_{t}$ governs the generation of her signal $\bm{\omega}^{k}_{t}=\{\omega^{k}_{i,t}\}_{i\in\mathcal{N}}$, the principal's choice of $\bm{\mathcal{P}}^{k}_{t}$ partially influences $\bm{\mathcal{P}}_{t}\equiv\{\bm{\mathcal{P}}^{k}_{t}, \bm{\mathcal{P}}^{-k}\}$ and the realizations of signal batch $\bm{W}_{t}=\{\bm{\omega}^{k}_{t}, \bm{W}^{-k}_{t}\}$.
%
%
This process is \textit{information design:}

\begin{definition}[\textbf{Information Design}]
An information design problem is defined as a tuple $\mathcal{I}\equiv<M[\bm{\theta}], \bm{\pi},$  $ \{\mathcal{P}^{k}_{i,t}\}_{i\in N, t\geq 0}, \Omega, \bm{\kappa}>$. 
Here, $M$ is an augmented Markov game model. $\bm{\pi}$ is the agents' policy profile.
$<\mathcal{P}^{k}_{i}, \Omega>$ is the information structure, where $\mathcal{P}^{k}_{i}\equiv\{\mathcal{P}^{k}_{i,t}\}_{t\geq 0}$ defines a distribution of the signal $\omega^{k}_{i,t}$ privately observed by agent $i$ at $t$.
$\bm{\kappa}(\cdot, \bm{\theta}): \mathcal{G} \mapsto \Delta(\mathcal{A}^{n})$ is the principal's goal, i.e., her target equilibrium probability distribution of agents' joint action conditioning only on the state and the agents' type.
\end{definition}

A solution to $\mathcal{I}$ is a stationary \textit{signaling rule profile} (signaling rule)
$\bm{\alpha}: \mathcal{G}\times \bm{\Theta}\mapsto \Delta(\Omega^{n})$ that defines $\bm{\mathcal{P}}^{k}\equiv\{\mathcal{P}^{k}_{i}\}_{i\in\mathcal{N}}$ of the joint signal $\bm{\omega}^{k}_{t}$. 
The signaling rule $\bm{\alpha}$ is \textit{correlated} if it is a joint function; i.e., $\bm{\alpha}(\bm{\omega}^{k}_{t}|g,\bm{\theta})\neq \prod_{i\in\mathcal{N}} \alpha_{i}(\omega^{k}_{i,t}|g,\bm{\theta})$, where $\alpha_{i}(\omega^{k}_{-i,t}|g,\bm{\theta})\equiv \sum_{\bm{\omega}^{k}_{-i,t}}\bm{\alpha}(\bm{\omega}^{k}_{t}|g,\bm{\theta})$.
The rule $\bm{\alpha}$ is \textit{independent} if the principal specifies the signal to each agent is independent of each other; i.e., $\bm{\alpha}(\bm{\omega}^{k}_{t}|g,\bm{\theta})= \prod_{i\in\mathcal{N}} \alpha_{i}(\omega^{k}_{i,t}|g,\bm{\theta})$.
Since the agents use Markov strategies, agent $i$'s period-$t$ action $a_{i,t}$ depends on histories $g^{t}$ and $\omega^{k;(t)}_{i}$ (via the selection rule $\beta_{i}$) only through the current-period $g_{t}$ and $\omega^{k}_{i,t}$.
Hence, we restrict attention to a Markovian signaling rule $\bm{\alpha}$ that specifies the distribution of period-$t$ signal by depending on the current state $g_{t}$ and the joint type $\bm{\theta}$.
We will denote the game $M$ with the principal using $\bm{\alpha}$ as $M[\bm{\alpha}|\bm{\theta}]$.

The information design problem is a planning problem. Hence, the design of $\bm{\alpha}$ is independent of any realizations of states.
Additionally, the principal does not know in advance all the possible equilibria that could be induced by any of her available signaling rules. 
Therefore, the principal's information design has to take into account how the signaling rule can induce the agents' behaviors that constitute an equilibrium.
If the information design is viewed as an extensive form game between the principal and the agents, the timing is as follows:
%
\begin{itemize}
    \item[(i)] The principal chooses a signaling rule profile $\bm{\alpha}$ for the agents.
    \item[(ii)] At the beginning of each period $t$, a state $g_{t}$ is realized and observed by the principal and all agents.
    %
    
    \item[(iii)] The principal sends a joint signal $\bm{\omega}^{k}_{t}$ according to $\bm{\alpha}$. Each agent $i$ receives $W_{i,t}=\{\omega^{k}_{i,t}, W^{-k}_{i,t}\}$ and observes $\bm{W}_{-i,t}\equiv \{\omega^{k}_{j,t}, W^{-k}_{j,t}\}_{j\neq i}$. 
    Here, $\bm{W}^{-k}_{t}\equiv\{W^{-k}_{i,t}, \bm{W}^{-k}_{-i,t}\}$ is generated according to $\big(\mathcal{P}^{-k}_{t}\big)^{n}$.
    %
    %
    \item[(iv)] The agents use $\bm{\beta}$ to select signals $\bm{\omega}_{t}$ from $\bm{W}_{t}$.
    %
    \item[(v)] Then, the agents use $\bm{\pi}$ to chooses their actions from $\mathcal{A}^{n}$ based on $g_{t}$ and $\bm{\omega}_{t}$.
    %
    %
    \item[(vi)] Immediate rewards are realized and the state $g_{t}$ is transitioned to $g_{t+1}$ according to $\mathcal{T}_{g}$.
\end{itemize}

\subsection{Equilibrium Concepts}

In this section, we define a stationary equilibrium concept of the game $M[\bm{\alpha}|\bm{\theta}]$.
With a slight abuse of notation, we suppress the notations of $\mathcal{P}^{-k}_{i}$, $W^{-k}_{i,t}$, and $\bm{W}^{-k}_{t}$ and only show $\mathcal{P}^{k}_{i,t}$, $\omega^{k}_{i,t}$ (of $W_{i,t}$), and $\bm{\omega}^{k}_{t}$ (of $\bm{W}_{t}$), for all $i\in\mathcal{N}$, $t\geq 0$, unless otherwise stated.
Since we focus on stationary environment, we suppress the time indexes from the notations, unless otherwise stated.

Similar to the canonical game $\widehat{M}$, Ionescu Tulcea theorem (see, e.g., \cite{hernandez2012discrete}) implies that the initial distribution $d_{g}$ of the state, the transition function $\mathcal{T}_{g}$, the distribution $\bm{\mathcal{P}}^{-k}$, the signaling rule $\bm{\alpha}$, and the strategy profile $<\bm{\beta}, \bm{\pi}>$ together define a unique probability measure $P^{\bm{\alpha}, \bm{\beta}}_{\bm{\pi}}$ on $(\mathcal{G} \times \Omega^{m\times n} \times\mathcal{A}^{n})^{\infty}$.
The expectation with respect to $P^{\bm{\alpha}, \bm{\beta}}_{\bm{\pi}}$ is denoted by $\mathbb{E}^{\bm{\beta},\bm{\alpha}}_{\bm{\pi}}\big[ \cdot \big]$ or $\mathbb{E}^{\bm{\beta},\bm{\alpha}}_{\bm{\pi}}\big[ \cdot \big| \cdot\big]$. 
%
With a slight abuse of notation, let $\mathcal{T}^{\bm{\alpha}, \bm{\beta}, \bm{\pi}}_{g'g}(\bm{\omega};\bm{\omega}^{k})$ denote the transition probability from state $g$ to state $g'$, given that the signal batch is $\bm{\omega}^{k}$ and the agents select $\bm{\omega} = \bm{\beta}(g, \bm{\theta}|\bm{\omega}^{k})$: for any $\bm{\omega}^{k}\in \Omega^{n}$ with $\bm{\alpha}(\bm{\omega}^{k}|g, \bm{\theta})>0$,
\begin{equation*}
    \begin{aligned}
    &\mathcal{T}^{\bm{\beta}, \bm{\pi} }_{g',g }(\bm{\omega};\bm{\omega}^{k})\equiv \sum_{\bm{a}}\bm{\pi}\big(\bm{a}|g, \bm{\omega}, \bm{\theta}\big) \mathcal{T}_{g}(g'|g, \bm{a}).
    \end{aligned}
\end{equation*}
%
%
Let $\mathcal{T}^{\bm{\alpha}, \bm{\beta}, \bm{\pi} }_{g',g }= \sum_{\bm{\omega}^{k}}\mathcal{T}^{ \bm{\beta}, \bm{\pi} }_{g',g }(\bm{\omega};$ $\bm{\omega}^{k})\bm{\alpha}(\omega^{k}|g,\bm{\theta})$.
%
%
Given $\bm{\alpha}$, $\bm{\beta}$, $\bm{\pi}$, define the \textit{state-signal value} function $V_{i}^{\bm{\pi}, \bm{\beta}, \bm{\alpha}}$ of agent $i$, representing agent $i$'s expected reward, originating at some $g, \bm{\omega}^{k}\in \mathcal{G}\times \Omega^{n}$ with $\bm{\alpha}(\bm{\omega}^{k}|g, \bm{\theta})>0$, agents select $\bm{\omega} = \bm{\beta}(g, \bm{\theta}|\bm{\omega}^{k})$,
%
%
\begin{equation}\label{eq:state_signal_value_function}
    \begin{aligned}
    &V^{\bm{\alpha}, \bm{\beta}, \bm{\pi} }_{i} (g, \bm{\omega}; \bm{\omega}^{k} |\bm{\theta}) \\
    &\equiv \sum_{t=0}^{\infty}\sum_{g'}
   \gamma^{t}\big(\mathcal{T}^{\bm{\beta},\bm{\pi} }_{g',g }(\bm{\omega}; \bm{\omega}^{k})\big)^{t}\sum_{\bm{a}', \bm{\omega}^{k'} }  \bm{\pi}(\bm{a}'|g', \bm{\omega}', \bm{\theta})\\
   &\times\bm{\alpha}(\bm{\omega}^{k'}|g', \bm{\theta})
  R_{i}(\bm{a}', g',\omega'_{i}|\theta_{i}) ,
    \end{aligned}
\end{equation}
where $\bm{\omega}' = \bm{\beta}(g, \bm{\theta}|\bm{\omega}^{k'})$.
%
%

Define the \textit{state value function} $J^{\bm{\pi}, \bm{\beta}, \bm{\alpha}}_{i}$ of agent $i$ that describes his expected reward, originating at any state $g\in \mathcal{G}$:
\begin{equation}\label{eq:state_value_function}
    \begin{aligned}
    &J^{\bm{\alpha}, \bm{\beta}, \bm{\pi}}_{i}(g|\bm{\theta})\equiv \sum_{t=0}^{\infty} \sum_{g'} \gamma^{t}\big(\mathcal{T}^{\bm{\alpha}, \bm{\beta}, \bm{\pi}}_{g',g}\big)^{t}\\
    &\times \sum_{\bm{a}', \bm{\omega}^{k'} }  \bm{\pi}(\bm{a}'|g', \bm{\beta}(g',  \bm{\theta}| \bm{\omega}^{k'}), \bm{\theta})\bm{\alpha}(\bm{\omega}^{k'}|g', \bm{\theta})\\
    &\times R_{i}(\bm{a'}, g', \beta_{i}(\bm{a}', g', \theta_{i}|\omega^{k'}_{i}) |\theta_{i}).
\end{aligned}
\end{equation}
%
Define the \textit{state-signal-action value function} $Q^{\bm{\pi}, \bm{\beta}, \bm{\alpha}}_{i}$ that represents agent $i$'s expected reward if $(\bm{\omega}, \bm{a})\in \Omega^{n} \times \mathcal{A}^{n}$ are played in $(g, \bm{\omega}^{k})\in \mathcal{G}\times \Omega^{n}$:
\begin{equation}\label{eq:agents_Q_function}
    \begin{aligned}
&Q^{\bm{\alpha}, \bm{\beta}, \bm{\pi} }_{i}  (\bm{a}, g, \omega_{i}  ;\omega^{k}_{i}|\bm{\theta})\equiv R_{i}(\bm{a},g,\omega_{i}|\theta_{i})\\
&+ \gamma \sum_{g'} \mathcal{T}_{g}(g'|g,\bm{a})\Big(\sum_{s=0}^{\infty}\sum_{g''} \gamma^{s}\big(\mathcal{T}^{\bm{\alpha},\bm{\beta}, \bm{\pi}}_{g'',g'}\big)^{s} \Big)\times\\
&\sum_{\bm{a}'', \bm{\omega}^{k''}} \bm{\pi}(\bm{a}''|g'', \bm{\omega}''; \bm{\theta})\bm{\alpha}(\bm{\omega}^{k''}|g'',\bm{\theta}) R_{i}(\bm{a}',g'', \omega'_{i}|\theta_{i})  \Big),
\end{aligned}
\end{equation}
where $\bm{\omega}'' = \bm{\beta}(g'', \bm{\theta}|\bm{\omega}^{k''})$.

We define an equilibrium concept known as \textit{sequential Markov perfect equilibrium} (SMPE) as follows.

\begin{definition}[SMPE]\label{def:PBE_markov}
Fix any signaling rule $\bm{\alpha}$. A strategy profile $<\bm{\beta}^{*}, \bm{\pi}^{*}>$ constitutes a \textit{(stationary) sequential Markov perfect equilibrium} (SMPE), where \textit{(i)} the policy profile is independent \big(i.e., $\bm{\pi}^{*}(\bm{a}|g,\bm{\omega}, \bm{\theta}) = \prod_{i\in\mathcal{N}} \pi^{*}_{i}(a_{i}|g, \omega_{i}, \theta_{i})$\big)
and \textit{(ii)} the agents are \textbf{sequential-perfectly rational} (sequential-perfect rationality), i.e., for any $g\in\mathcal{G}$, $\vec{\beta}'_{i}\equiv\{\beta'_{i,\tau}\}_{\tau\geq 0}$, $i\in\mathcal{N}$,
\begin{equation}\label{eq:SPR_beta}
    \begin{split}
        J^{\bm{\alpha},\bm{\beta}^{*}, \bm{\pi}^{*}}_{i} (g |\bm{\theta}) \geq J^{\bm{\alpha},\vec{\beta}'_{i},\bm{\beta}^{*}_{-i}, \bm{\pi}^{*}}_{i} (g |\bm{\theta}),
    \end{split}
\end{equation}
and, for any $\omega^{k}_{i}\in \Omega$ with $\alpha_{i}(\omega^{k}_{i}|g, \bm{\theta})>0$, $\omega^{*}_{i} = \beta^{*}_{i}(g, \theta_{i}|\omega^{k}_{i})$, $\vec{\pi}'_{i}\equiv\{\pi'_{i,\tau}\}_{\tau\geq 0}$,
\begin{equation}\label{eq:SPR_pi}
    \begin{aligned}
    &\mathbb{E}^{\bm{\beta}^{*}}_{\bm{\omega}^{k}_{-i}\sim\bm{\alpha}_{-i}}\Big[ V^{\bm{\alpha}, \bm{\beta}^{*}, \bm{\pi}^{*} }_{i}(g, \omega^{*}_{i}, \bm{\omega}^{*}_{-i}; \omega^{k}_{i}, \bm{\omega}^{k}_{-i}|\bm{\theta}) \Big]\\
    &\geq \mathbb{E}^{\bm{\beta}^{*}}_{\bm{\omega}^{k}_{-i}\sim\bm{\alpha}_{-i}}\Big[ V^{\bm{\alpha},\bm{\beta}^{*}, \vec{\pi}'_{i},\bm{\pi}^{*}_{-i} }_{i} (g, \omega^{*}_{i}, \bm{\omega}^{*}_{-i}; \omega^{k}_{i}, \bm{\omega}^{k}_{-i}|\bm{\theta}) \Big].
    \end{aligned}
\end{equation}
\end{definition}

An SMPE extends the stationary Markov perfect equilibrium (see, e.g., \cite{he2017stationary}) to our augmented Markov game $M[\bm{\alpha}|\bm{\theta}]$
The sequential-perfect rationality describes the coupled sequential best responses of each agent's selection and action given a state and the available signal batch.
In words, a strategy profile $<\bm{\beta}^{*}, \bm{\pi}^{*}>$ is sequential-perfectly rational if \textit{(i)} given that agents choose actions according to the equilibrium policy profile $\bm{\pi}^{*}$, there is \textit{no} state $g\in \mathcal{G}$ such that once it is reached, the agents strictly prefer to deviate from $\bm{\beta}^{*}$; and \textit{(ii)} there is \textit{no} information set $(g, \{\bm{\omega}^{*}, \bm{W}^{-k}\})\in \mathcal{G}\times \Omega^{m\times n}$ where $\bm{\omega}^{*}$ is selected by $\bm{\beta}^{*}$ such that once it is reached, the agents strictly prefer to deviate from $\bm{\pi}^{*}$.

The concept of SMPE in Definition \ref{def:PBE_markov}, however, permits arbitrarily complex and possibly nonstationary deviations from (stationary) equilibrium profile $<\bm{\beta}^{*}, \bm{\pi}^{*}>$. 
The following lemma states that it entails no loss of generality to consider any one-shot deviations from $<\bm{\beta}^{*}, \bm{\pi}^{*}>$.

\begin{lemma}\label{lemma:one-shot-deviation}
Let $\vec{\beta}'_{i}\circ 0\equiv\{\beta'_{i, \tau}\}_{\tau\geq 0}$ and $\vec{\pi}'_{i}\circ 0 \equiv \{\pi'_{i,\tau}\}_{\tau\geq 0}$, respectively, be such that $\beta'_{i,\tau} = \beta^{*}_{i}$ and $\pi'_{i,\tau} = \pi^{*}_{i}$, for all $\tau\geq 1$, while $\beta'_{i,0}$ and $\pi'_{i,0}$ are any two arbitrary strategies.
A strategy profile $<\bm{\beta}^{*}, \bm{\pi}^{*}>$ constitutes a sequential-perfectly rational equilibrium profile of an SMPE if and only if 
for any $g_{0}\in \mathcal{G}$, $i\in\mathcal{N}$,
\begin{equation}\label{eq:SPR_beta_1shot}
    \begin{split}
        J^{\bm{\beta}^{*}, \bm{\pi}^{*}, \bm{\alpha}}_{i} (g_{0} |\theta_{i}) \geq J^{\vec{\beta}'_{i}\circ 0,\bm{\beta}^{*}_{-i}, \bm{\pi}^{*}, \bm{\alpha}}_{i} (g_{0} |\theta_{i}),
    \end{split}
\end{equation}
and, for any $\omega^{k}_{i}\in\Omega$ with $\alpha_{i}(\omega^{k}_{i}|g_{0}, \bm{\theta})>0$, $\omega^{*}_{i} = \beta^{*}_{i}(g_{0},  \theta_{i}|\omega^{k}_{i})$, $i\in\mathcal{N}$, $\pi'_{i,0}$,
\begin{equation}\label{eq:SPR_pi_1shot}
    \begin{split}
        V^{\bm{\beta}^{*}, \bm{\pi}^{*}, \bm{\alpha}}_{i}& (g_{0}, \omega^{*}_{i,0}; \omega^{k}_{i,0}|\theta_{i})\\
        \geq & V^{\bm{\beta}^{*}, \vec{\pi}'_{i}\circ 0,\bm{\pi}^{*}_{-i}, \bm{\alpha}; \mu_{i}}_{i} (g_{0}, \omega^{*}_{i,0}; \omega^{k}_{i,0}|\theta_{i}).
    \end{split}
\end{equation}
\end{lemma}

A one-shot deviation is a behavior of each agent $i$'s deviating from the equilibrium profile $<\bm{\beta}^{*}, \bm{\pi}^{*}>$ by selecting a signal using any $\beta'_{i,0}$ and taking an action using any $\pi'_{i,0}$ at the initial period of any subgame of $M[\bm{\alpha}|\bm{\theta}]$, then reverting back to his equilibrium profile for the rest of the game.
The one-shot deviation property in Lemma \ref{lemma:one-shot-deviation} allows the principal to restrict attention to the equilibrium characterization by considering the robustness of the information design to the one-shot deviation without lose of generality.

\section{SMPE Implementability}\label{sec:infoD_Myersonian}

In this section, we formally formulate the principal's information design problem.
The optimality criterion of successful information design is captured by the notion of \textit{implementability} which is characterized in the equilibrium concept of SMPE.

\subsection{Implementability}

As in a canonical Markov game, each agent $i$'s decision of choosing an action $a$ takes into account other agents' decisions of choosing $\bm{a}_{-i}$ because its immediate reward of taking $a_{i}$ directly depends on $\bm{a}_{-i}$.
In an augmented Markov game $M[\bm{\alpha}|\bm{\theta}]$, agent $i$'s choices of $\beta_{i}$ and $\pi_{i}$ are coupled because $\omega_{i}$ specified by $\beta_{i}$ has a direct causal effect on $a_{i}$ through $\pi_{i}$.
Thus, each agent's immediate reward indirectly depends on other agents' selected signals through their actions.
Hence, agents' selection of signals is also a part of the strategic interactions in a $M[\bm{\alpha}|\bm{\theta}]$.
Since $\bm{\mathcal{P}}^{-k}$ is fixed, the principal's choice of $\bm{\alpha}$ controls the dynamics of $\bm{W}_{t}$ given the strategy profiles.
Therefore, it is possible for the principal to influence the equilibrium behaviors of agents in $M[\bm{\alpha}|\bm{\theta}]$ through proper designs of $\bm{\alpha}$.

The principal's information design takes an objective-first approach to design the information structure (given $\Omega$) of the signals sent to agents, toward desired objectives $\bm{\kappa}$, in a strategic setting through the design of $\bm{\alpha}$, where self-interested agents act rationally by choosing $\bm{\beta}$ and $\bm{\pi}$.
Although any realization of the signal depends on the current state, the choice of $\bm{\alpha}$ is independent of the realizations of the states.
%
%
The key restriction on the principal's $\bm{\alpha}$ is that the agents are elicited to perform equilibrium actions that coincide with the principal's goal, which is referred to as \textit{admissibility.}

\begin{definition}[Admissibility]\label{def:admissibility}
Fix any $\bm{\kappa}$ and $\bm{\alpha}$. 
Let $<\bm{\beta}^{*}, \bm{\pi}^{*}>$ be any SMPE of the game $M[\bm{\alpha}|\bm{\theta}]$.
The policy profile $\bm{\pi}^{*}$ is \textbf{admissible} if, for all $g\in\mathcal{G}$,
\begin{equation}\label{eq:def_admissible}
    \begin{aligned}
    &\bm{\kappa}(\bm{a} | g, \bm{\theta})=\sum_{ \bm{\omega}^{k} }\bm{\pi}^{*}\big(\bm{a}|g, \bm{\beta}^{*}(g, \bm{\theta}|\bm{\omega}^{k}), \bm{\theta}\big)\bm{\alpha}(\bm{\omega}^{k}|g, \bm{\theta}).
       \end{aligned}
\end{equation}
\end{definition}

The admissibility imposes a constraint on the signaling rule $\bm{\alpha}$ and the induced policy profile $\bm{\pi}$ such that the goal $\bm{\kappa}$ is achieved in the sense of (\ref{eq:def_admissible}).
%
%
We define a strong version of admissibility as follows.

\begin{definition}[Strong Admissibility]\label{def:strong_admissible}
Fix any $\bm{\kappa}$ and $\bm{\alpha}$. 
Let $<\bm{\beta}^{*}, \bm{\pi}^{*}>$ be any SMPE of the game $M[\bm{\alpha}|\bm{\theta}]$.
The policy profile $\bm{\pi}^{*}$ is \textbf{string admissible} if, for all $g\in\mathcal{G}$, $\omega^{k}_{i}\in\Omega$, $i\in\mathcal{N}$,
\begin{equation}\label{eq:def_strongly_admissible}
    \begin{aligned}
    &\bm{\kappa}(\bm{a}|g, \bm{\theta})\sum\limits_{\bm{\omega}^{k}_{-i} }\bm{\alpha}(\omega_{i}, \bm{\omega}^{k}_{-i}|g,\bm{\theta})\bm{\beta}^{*}(g, \bm{\theta}|\omega^{k}_{i},\bm{\omega}^{k}_{-i})\\
    &= \sum\limits_{\bm{\omega}^{k}_{-i} } \bm{\pi}^{*}\big(\bm{a}|g, \bm{\beta}^{*}(g, \bm{\theta}|\omega^{k}_{i},\bm{\omega}^{k}_{-i}), \bm{\theta}\big)\bm{\alpha}(\bm{\omega}^{k}|g, \bm{\theta}).
    \end{aligned}
    \end{equation}
\end{definition}

The strong version constrains $\bm{\alpha}$ and $\bm{\pi}$ additionally by imposing individual-level constraint on the signaling rule.
It is straightforward to verify that the strong admissibility implies admissibility but not vice versa.

The success criterion of information design for the game $M[\bm{\alpha}|\bm{\theta}]$ in equilibrium is captured by the notion of \textit{implementability}.

\begin{definition}[SMPE Implementability]\label{def:implementability}
Given any $\bm{\kappa}$. We say that the signaling rule $\bm{\alpha}$ is (strongly) \textbf{implementable} in SMPE ( SMPE Implementability) if $M[\bm{\alpha}|\bm{\theta}]$ has an SMPE $<\bm{\beta}^{*}, \bm{\pi}^{*}>$ in which $\bm{\pi}^{*}$ is (strongly) admissible. 
%
%
\end{definition}

The SMPE implementability requires that \textit{(i)} the signaling rule $\bm{\alpha}$ designed by the principal induces a SMPE of $M[\bm{\alpha}|\bm{\theta}]$ and \textit{(ii)} the principal's goal is achieved (i.e., the equilibrium policy profile is admissible or strong admissible).
%
%

Given any $\bm{\mathcal{P}}^{-k}$, the distribution of the action $\bm{a}$ conditioning on any state $g$ is jointly determined by the agents' $\bm{\beta}$ and the principal's $\bm{\alpha}$.
Hence, given $\bm{\mathcal{P}}^{-k}$, the signal $\bm{\omega}^{k}$ sent by the principal by using $\bm{\alpha}$ ultimately influences each agent' expected reward. 
However, this information is transmitted \textit{indirectly} through the agents' selection rules when the selected signal $\bm{\omega} = \bm{\beta}(g,\bm{\theta}|\{\bm{\omega}^{k}, \bm{W}^{-k}\})$ is not equal to $\bm{\omega}^{-k}$.
We refer to the design of $\bm{\alpha}$ that induces such selection rules in SMPE of $M[\bm{\alpha}|\bm{\theta}]$ as an \textit{indirect information design} (IID).
We call the game with such $\bm{\alpha}$ as \textit{indirect augmented Markov game}, denoted by $M^{-D}[\bm{\alpha}]$.

\subsection{Direct Information Design}\label{sec:direct_IDP}

As a designer, the principal takes into consideration how each agent strategically behaves according to the game rules and reactions to his opponents' behaviors as well as the responses from the environment.
In any $M^{-D}[\bm{\alpha}]$, the principal's design of $\bm{\alpha}$ must predict the agents' equilibrium selection rule profile $\bm{\beta}$ and their corresponding equilibrium policy profile $\bm{\pi}$ that might be induced by $\bm{\alpha}$.
In contrast to $M^{-D}[\bm{\alpha}]$, the principal may elect to a \textit{direct information design} in which the principal makes her signals payoff-relevant to each agent's decision of choosing an action by incentivizing each agent to select her signal at each state.
We refer to the game with such signaling rule as \textit{direct augmented Markov game}, denoted by $M^{D}[\bm{\alpha}]$.
The implementability of the direct information design requires a restriction noted as \textit{obedience} in addition to the admissibility.

\begin{definition}[Obedience]\label{def:obedience}
In any $M^{D}[\bm{\alpha}]$, agent $i$'s selection rule $\beta_{i}$ is \textbf{dominant-strategy obedient} (DS-obedient, DS-obedience) if, for any $(g,  \{\omega^{k}_{i}, W^{-k}_{i}\})$  $\in \mathcal{G}\times \Omega^{m}$ with $\alpha_{i}(\omega^{k}_{i}|g, \bm{\theta})>0$, $i\in\mathcal{N}$,
\begin{equation}
        \begin{split}
            \beta_{i}(g_{t}, \theta_{i}|\{\omega^{k}_{i,t}, W^{-k}_{i,t}\}) = \omega^{k}_{i,t}.
        \end{split}
\end{equation}
Agent $i$'s selection rule $\beta_{i}$ is \textbf{Bayesian obedient} (Bayesian obedience) if, for any $(g,  \{\omega^{k}_{i}, W^{-k}_{i}\})\in $  $\mathcal{G}\times \Omega^{m}$ with $\alpha_{i}(\omega^{k}_{i}|g, \bm{\theta})>0$, $i\in\mathcal{N}$,
\begin{equation}
        \begin{split}
            \beta_{i}(g_{t}, \theta_{i}|\{\omega^{k}_{i,t}, W^{-k}_{i,t}\}) = \omega^{k}_{i,t},
        \end{split}
\end{equation}
\textbf{when all other agents are obedient,} i.e., $\bm{\beta}_{-i}(g, \bm{\theta}_{-i}|$  $\bm{\omega}^{k}_{-i}) = \bm{\omega}^{k}_{-i}$.
\end{definition}

We will refer to a SMPE with obedient selection rule profile as (dominant-strategy or Bayesian) obedient SMPE (O-SMPE).

\begin{definition}[OIL]\label{def:O_Implementability}
Given any goal $\bm{\kappa}$, the signaling rule $\bm{\alpha}$ is (DS, Bayesian) \textbf{obedient-implementable} in SMPE (OIL) if it induces an (DS, Bayesian) O-SMPE $<\bm{\beta}, \bm{\pi}>$ in which $\bm{\beta}$ is (DS, Bayesian) obedient and $\bm{\pi}$ is admissible. 
%
%
\end{definition}

In a $M^{D}[\bm{\alpha}|\bm{\theta}]$, the principal wants $\bm{\omega}^{k}_{t}$ to directly enter the immediate rewards of the agents at each period $t$. 
The OIL guarantees that \textit{(i)} agents would want to select the signal $\bm{\omega}^{k}$ sent by the principal than choose any other signals from $\bm{W}^{-k}$, and \textit{(ii)} agents take actions specified by the admissible policy other than other available actions.
Hereafter, we denote obedient selection rule by $\bm{\beta}^{O}$ and $\beta^{O}_{i}$, for all $i\in\mathcal{N}$.

A successful information design depends on the principal's having accurate beliefs in regard to the agents' decision processes. This includes all the possible indirect selection behaviors of the agents, i.e., all possible $\bm{\beta}\neq \bm{\beta}^{O}$. 
The point of direct information design is that it allows the principal to ignore analyzing all of agents' indirect selections behaviors and focus on the obedient $\bm{\beta}^{O}$.
%

\subsection{Characterizing OIL}

In this section, we characterize the OIL and formulate the principal's information design problem given a goal $\bm{\kappa}$.

The following proposition is an analog of Bellman's Theorem \cite{bellman1966dynamic}.
\begin{proposition}\label{prop:Bellman_theorem}
Given a game $M[\bm{\alpha}|\bm{\theta}]$, for any stationary $<\bm{\beta}, \bm{\pi}>$, any $V_{i}: \mathcal{G} \times \Omega \times \Omega \mapsto \mathbb{R}$, any $J_{i}: \mathcal{G} \mapsto \mathbb{R}$, any $Q_{i}: \mathcal{G} \times \Omega \times \mathcal{A}^{n} \times \Omega \mapsto \mathbb{R}$, we say $V_{i} = V^{\bm{\alpha}, \bm{\beta}, \bm{\pi}}_{i}$ in (\ref{eq:state_signal_value_function}), $J_{i}=J^{\bm{\alpha}, \bm{\beta}, \bm{\pi}}_{i}$ in (\ref{eq:state_value_function}), and $Q_{i} = Q^{\bm{\alpha}, \bm{\beta}, \bm{\pi}}_{i}$ in (\ref{eq:agents_Q_function}), if and only if the following unique Bellman recursions are satisfied:
\begin{equation}\label{eq:bellman_V_1}
    \begin{aligned}
    &V_{i}(g, \bm{\omega}; \bm{\omega}^{k}|\bm{\theta}) = \sum_{\bm{a}} \bm{\pi}(\bm{a}|g, \bm{\omega}, \bm{\theta})Q_{i}  (\bm{a},g, \omega_{i};\omega^{k}_{i}|\bm{\theta}),
    \end{aligned}
\end{equation}
\begin{equation}\label{eq:bellman_J_1}
    \begin{aligned}
    &J_{i}(g|\bm{\theta}) = \sum_{ \bm{\omega}^{k} }\bm{\alpha}(\bm{\omega}^{k}|g, \bm{\theta}) V_{i}(g, \bm{\beta}(g,  \bm{\theta}|\bm{\omega}^{k}); \bm{\omega}^{k}|\bm{\theta}),
    \end{aligned}
\end{equation}
\begin{equation}\label{eq:bellman_Q_1}
    \begin{aligned}
    &Q_{i}(\bm{a}, g, \omega_{i}; \omega^{k}_{i}|\bm{\theta})\\
    &= R_{i}(\bm{a}, g, \omega_{i} |\theta_{i}) + \gamma \sum_{g'} \mathcal{T}_{g}(g'|g, \bm{a})J_{i}(g'|\bm{\theta}).
    \end{aligned}
\end{equation}
\end{proposition}

From Proposition \ref{prop:Bellman_theorem}, we can re-define $V^{\bm{\alpha}, \bm{\beta}, \bm{\pi}}_{i}$, $J^{\bm{\alpha}, \bm{\beta}, \bm{\pi}}_{i}$, and $Q^{\bm{\alpha}, \bm{\beta}, \bm{\pi}}_{i}$ given in (\ref{eq:state_signal_value_function})-(\ref{eq:agents_Q_function}), respectively, as the unique pair of value functions that satisfy the Bellman recursions (\ref{eq:bellman_V_1}), (\ref{eq:bellman_J_1}), and (\ref{eq:bellman_Q_1}).

\begin{lemma}\label{lemma:Bellman_recursive_optimality}
Fix $\bm{\alpha}$.
Let $\bm{\beta}^{O} = \{\beta^{O}_{i}\}_{i\in\mathcal{N}}$ denote the obedient selection rule profile.
The strategy profile $<\bm{\beta}^{O},  \bm{\pi}^{*}>$ is a (DS, Bayesian) O-SMPE if and only if, for any $g\in\mathcal{G}$, $\omega^{k}_{i}\in \Omega$ with $\alpha_{i}(\omega^{k}_{i}|g, \bm{\theta})>0$, $i\in\mathcal{N}$,
\begin{equation}\label{eq:Bellma_optimality_V}
    \begin{aligned}
    &V^{\bm{\alpha},  \bm{\beta}^{O},\bm{\pi}^{*} }_{i}(g,\bm{\omega}^{k}|\bm{\theta}) \\
    &= \max_{a'_{i}} \mathbb{E}_{\bm{a}_{-i}\sim\bm{\pi}^{*}_{-i}}\Big[Q^{\bm{\beta}^{O}, \bm{\alpha};\mu_{i}}(a'_{i}, \bm{a}_{-i},g;\omega^{k}_{i}|\theta_{i})\Big],
    \end{aligned}
\end{equation}
\begin{itemize}
    \item[(i)] and $\bm{\beta}^{O}$ is DS-obedient, i.e., if, for any arbitrary selection rule profiles $\bm{\hat{\beta}}_{-i}$, any $g\in\mathcal{G}$, $i\in\mathcal{N}$,
    \begin{equation}\label{eq:Bellma_optimality_J_DS}
    \begin{aligned}
    &J^{\bm{\alpha},\beta^{O}_{i}, \bm{\hat{\beta}}_{-i}, \bm{\pi}^{*} }_{i}(g|\bm{\theta}) = \max_{\omega'_{i}}\sum_{ \bm{\omega}^{k}_{-i}}\bm{\alpha}_{-i}(\bm{\omega}^{k}_{-i}|g, \bm{\theta})\\
    &\times V^{\bm{\alpha},  \bm{\beta}^{O},\bm{\pi}^{*} }_{i}(g, \omega'_{i}, \bm{\hat{\beta}}_{-i}(g_{t}, \bm{\theta}_{-i}|\bm{\omega}^{k}_{-i}); \bm{\omega}^{k}|\bm{\theta});
    \end{aligned}
\end{equation}
%
\item[(ii)] or, $\bm{\beta}^{O}$ is Bayesian-obedient, i.e., if, for any $g\in \mathcal{G}$, $i\in\mathcal{N}$,
\begin{equation}\label{eq:Bellma_optimality_J_Bayesian}
    \begin{aligned}
    &J^{\bm{\alpha},\bm{\beta}^{*}, \bm{\pi}^{*} }_{i}(g|\bm{\theta}) \\
    &= \max_{\omega'_{i}}\sum_{ \bm{\omega}^{k}_{-i}}\bm{\alpha}_{-i}(\bm{\omega}^{k}_{-i}|g, \bm{\theta}) V^{\bm{\alpha},  \bm{\beta}^{O},\bm{\pi}^{*} }_{i}(g, \omega'_{i}, \bm{\omega}^{k}_{-i}; \bm{\omega}^{k}|\bm{\theta}).
    \end{aligned}
\end{equation}
\end{itemize}
\end{lemma}

In Lemma \ref{lemma:Bellman_recursive_optimality}, (\ref{eq:Bellma_optimality_V})-(\ref{eq:Bellma_optimality_J_DS}) \big(resp. (\ref{eq:Bellma_optimality_V})-(\ref{eq:Bellma_optimality_J_Bayesian})\big) constitute a recursive representation of a DS-O-SMPE \big(resp. Bayesian O-SMPE\big). 
Jointly, (\ref{eq:Bellma_optimality_V})-(\ref{eq:Bellma_optimality_J_DS}) require that \textit{(i)} the expected payoff of each agent $i$ is maximized by his equilibrium policy $\pi^{*}_{i}$ in every state and every signal selected according to any possible equilibrium selection rule $\bar{\beta}_{i}$, \textit{(ii)} the expected payoff of each agent $i$ is maximized by the obedient selection rule $\beta^{*}_{i}$ in every state when the action is taken by any possible corresponding equilibrium policy $\bar{\pi}_{i}$, and \textit{(iii)} $\bar{\beta}_{i} = \beta^{*}_{i}$ and $\bar{\pi}_{i} = \pi^{*}_{i}$ (given that his opponents are using any arbitrary selection rule but following O-SMPE policy profile.) Similar interpretation can be done for Bayesian O-SMPE, given that each agent $i$'s opponents are following Bayesian O-SMPE strategy profile.

\subsubsection{Design Regime: Fixed-Point Alignment}\label{sec:design_regime}

In this section, we restrict attention to the Bayesian obedience and formulate an information design regime for a given goal.
The following theorem states the existence of Bayesian O-SMPE.

\begin{theorem}\label{theorem:existence_SMPE}
Every augmented Markov game $M[\bm{\alpha}|\bm{\theta}]$ admits a stationary Bayesian O-SMPE for any regular $\bm{\alpha}$.
\end{theorem}

We provide an approach which we refer to as the \textit{fixed-point alignment} to formulate the design of the Bayesian OIL signaling rule $\bm{\alpha}$ as a planning problem.
First, we define a class of principal's goal which we refer to as \textit{Markov perfect goal}.
Let, for all $i\in\mathcal{N}$, 
\begin{equation*}
    \begin{aligned}
    R^{\bm{\alpha}}_{i}(\bm{a},  g|\theta_{i}) \equiv \sum\limits_{\omega_{i}^{k} } R_{i}(\bm{a}, g, \omega^{k}_{i}|\theta_{i})\alpha_{i}(\omega^{k}_{i}|g,\bm{\theta}).
    \end{aligned}
\end{equation*}

\begin{definition}[Markov Perfect Goal]
We say that the principal's goal $\bm{\kappa}$ is a Markov perfect goal if it is a MPE of a game canonical Markov game with $R^{\bm{\alpha}}_{i}$ as the reward function of each agent $i$; i.e., for all $\bm{a}_{t}=\{a_{i,t}, \bm{a}_{-i,t}\}\in\mathcal{A}^{n}$ with $\bm{\kappa}(\bm{a}_{t}|g_{t})>0$, $a'_{i,t}\in\mathcal{A}$, $g_{t}\in \mathcal{G}$, $t\geq 0$, $i\in\mathcal{N}$,
\begin{equation*}
    \begin{aligned}
    \mathbb{E}_{\bm{\kappa}_{-i}}\Big[\mathtt{Expr}_{i}&(g_{t},a_{i,t},\bm{a}_{-i,t};\theta_{i} |\bm{\kappa}, R^{\bm{\alpha}}_{i})\Big]\\
    &\geq \mathbb{E}_{\bm{\kappa}_{-i}}\Big[\mathtt{Expr}_{i}(g_{t},a'_{i,t},\bm{a}_{-i,t};\theta_{i} |\bm{\kappa}, R^{\bm{\alpha}}_{i})\Big],
    \end{aligned}
\end{equation*}
where $\mathtt{Expr}_{i}$ is defined in (\ref{eq:agent_expected_payoff_preliminary}).
Let $\mathtt{MPG}[\bm{R}]$ denote a set of Markov perfect goals given the reward functions $\bm{R}\equiv\{R_{i}\}_{i\in\mathcal{N}}$.
\end{definition}


For any state value function $J_{i}$, any $\bm{a}\in\mathcal{A}^{n}$, $g\in\mathcal{G}$, $\omega_{i},\omega^{k}_{i}\in\Omega$, we denote $Q_{i}(\bm{a}, g, \omega_{i}; \omega^{k}_{i}|\bm{\theta}; J_{i})$ as the state-signal-action value function constructed in terms of $J_{i}$ according to (\ref{eq:bellman_Q_1}).
Similarly, we denote $J_{i}(g|\bm{\theta}; V_{i})$ as the state value function constructed in terms of $V_{i}$ according to (\ref{eq:bellman_J_1}).
Given any policy profile $\bm{\pi}_{-i}$ of other agents and any state value function $J_{i}$, we let, for all $a_{i}\in\mathcal{A}$, $g\in\mathcal{G}$, $\bm{\omega}^{k}=\{\omega^{k}_{i}, \bm{\omega}^{k}_{-i}\}\in\Omega^{n}$, $\bm{\theta}=\{\theta_{i}, \bm{\theta}_{-i} \}\in\Theta^{n}$, $i\in\mathcal{N}$,
\begin{equation}\label{eq:pi_specified_Q}
    \begin{aligned}
    &Q^{\bm{\pi}_{-i}}_{i}(a_{i}, g, \omega_{i}, \bm{\omega}^{k}_{-i};\bm{\omega}^{k}|\bm{\theta};J_{i})\\
    &\equiv \mathbb{E}_{\bm{a}_{-i}\sim \bm{\pi}_{-i}(\cdot|g, \bm{\omega}^{k}_{-i}, \bm{\theta}_{-i})}\Big[ Q_{i}(a_{i},\bm{a}_{-i}, g, \omega_{i}; \omega^{k}_{i}|\bm{\theta}, J_{i}) \Big].
    \end{aligned}
\end{equation}
Similarly, for any state-signal value function $V_{i}$, we denote $Q^{\bm{\alpha}}_{i}(\cdot; V_{i})$ as the value function constructed in terms of $V_{i}$ according to the Bellman recursions (\ref{eq:bellman_V_1})-(\ref{eq:bellman_Q_1}), i.e., for all $\bm{a}\in\mathcal{A}^{n}$, $g\in\mathcal{G}$, $\omega^{k}_{i}\in \Omega$ with $\alpha_{i}(\omega^{k}_{i}|g, \bm{\theta})>0$, $i\in\mathcal{N}$, 
\begin{equation*}
    \begin{aligned}
    &Q^{\bm{\alpha}}_{i}(\bm{a}, g; \omega^{k}_{i}|\bm{\theta}; V_{i})= R_{i}(\bm{a}, g,\omega^{k}_{i}|\theta_{i}) \\
    &+ \gamma \sum_{g', \bm{\omega}^{k'}} \mathcal{T}_{g}(g'|g, \bm{a})\bm{\alpha}(\bm{\omega}^{k'}|g', \bm{\theta})V_{i}(g'; \bm{\omega}^{k'}|\bm{\theta}).
\end{aligned}
\end{equation*}

The Bayesian OIL restricts the design of the signaling rule $\bm{\alpha}$ by threefold constraints.
First, agents are incentivized to be Bayesian obedient in the signal selections. 
Second, the Bayesian-obedient agents' policy profile constitute the policy component of a Bayesian O-SMPE. Third, the equilibrium policy profile is admissible.
The first two constraints require the $\bm{\alpha}$ to elicit the agents to a Bayesian O-SMPE.

Proposition \ref{prop:Bayesian_OIL_equilibrium_condition} provides a general formulation of an optimization problem to find the policy profile of a Bayesian O-SMPE of a game $M[\bm{\alpha}|\bm{\theta}]$.

\begin{proposition}\label{prop:Bayesian_OIL_equilibrium_condition}
Fix a signaling rule $\bm{\alpha}$. Let $\bm{\beta}^{O}$ denote the Bayesian obedient selection rule profile. Let $\bm{V}^{*}=\{V^{*}_{i}\}_{i\in\mathcal{N}}$ denote the corresponding state-signal value function of a policy profile $\bm{\pi}^{*}$. 
The strategy profile $<\bm{\beta}^{O}, \bm{\pi}^{*}>$ constitutes a Bayesian O-SMPE if and only if $<\bm{\pi}^{*}, \bm{V}^{*}>$ is a solution of the following optimization problem with $\bm{Z}(\bm{\pi}^{*}, \bm{V}^{*}; \bm{\alpha}) = 0$:
\begin{equation}\tag{$\mathtt{Opt}$}\label{eq:nonlinear_program_pi_object_Bayesian}
    \begin{aligned}
    &\min\limits_{\bm{\pi}, \bm{V}} \bm{Z}(\bm{\pi}, \bm{V}; \bm{\alpha}) \\
    &\equiv \sum_{i, g, \bm{\omega}^{k}} V_{i}(g; \bm{\omega}^{k}|\theta_{i}) 
    -  \mathbb{E}_{\bm{a}\sim \bm{\pi}(\cdot|\bm{\omega}^{k})}\Big[Q^{\bm{\alpha}}_{i}(\bm{a}, g; \omega^{k}_{i}|\bm{\theta}; V_{i})\Big],
    \end{aligned}
\end{equation}
subject to, for all $i\in\mathcal{N}$, $g\in\mathcal{G}$, $\bm{\omega}^{k}\in \Omega^{n}$ with $\bm{\alpha}(\bm{\omega}^{k}|g, \bm{\theta})>0$, $a'_{i}\in \mathcal{A}$, $\omega'_{i}\in\Omega$, %
\begin{equation}\tag{$\mathtt{RG}_{i}$}\label{eq:regular_policy}
    \begin{aligned}
    \pi_{i}(a_{i}|g,\omega_{i}, \theta_{i})\geq 0, \sum_{a_{i}\in\mathcal{A}} \pi_{i}(a_{i}|g,\omega_{i}, \theta_{i}) =1,
    \end{aligned}
\end{equation}
\begin{equation}\tag{$\mathtt{FE}_{i}$}\label{eq:nonlinear_program_pi_constraint}
    \begin{split}
     &V_{i}(g;\bm{\omega}^{k}|\bm{\theta})\\
        &\geq \mathbb{E}_{\bm{a}_{-i}\sim \bm{\pi}_{-i}(\cdot|\bm{\omega}^{k}_{-i})}\Big[ Q^{\bm{\alpha}}_{i}(a'_{i},\bm{a}_{-i}, g; \omega^{k}_{i}|\bm{\theta}; V_{i})\Big],
    \end{split}
\end{equation}
\begin{equation}\tag{$\mathtt{BOB}0_{i}$}\label{eq:constraint_BOB_pi_V}
    \begin{aligned}
    &J^{\bm{\alpha},\bm{\beta}^{O}, \bm{\pi} }_{i}(g|\bm{\theta}; V_{i}) \\
    &\geq \sum_{ \bm{\omega}^{k}_{-i}}\bm{\alpha}_{-i}(\bm{\omega}^{k}_{-i}|g, \bm{\theta}) V_{i}(g, \omega'_{i}, \bm{\omega}^{k}_{-i}; \bm{\omega}^{k}|\bm{\theta}).
    \end{aligned}
\end{equation}
%
%
\end{proposition}

Proposition (\ref{prop:Bayesian_OIL_equilibrium_condition}) extends the fundamental formulation of finding the Nash equilibrium if a stochastic game as a nonlinear programming (Theorem 3.8.2 of \cite{filar1997competitive}; see also, \cite{prasad2012general,prasad2015two}).
Here, the condition (\ref{eq:regular_policy}) ensures that each $\pi_{i}$ is valid policy and rules out the possible trivial solution $\pi_{i}=0$ for all $i\in\mathcal{N}$.
The constraints (\ref{eq:nonlinear_program_pi_constraint}) and (\ref{eq:constraint_BOB_pi_V}) are two necessary conditions for a Bayesian O-SMPE of the game $M[\bm{\alpha}|\bm{\theta}]$ derived from (\ref{eq:Bellma_optimality_V}) and (\ref{eq:Bellma_optimality_J_Bayesian}) of Lemma \ref{lemma:Bellman_recursive_optimality}.
Any feasible solution $<\bm{\pi}, \bm{V}>$ making $\bm{Z}(\bm{\pi}, \bm{V}; \bm{\alpha}) = 0$ constitutes a Bayesian O-SMPE (in which the admissibility is not constrained).
Here, the Bayesian obedient selection rule profile $\bm{\beta}^{O}$ is not a solution of the optimization problem (\ref{eq:nonlinear_program_pi_object_Bayesian})-(\ref{eq:constraint_BOB_pi_V}); instead, the optimality of Bayesian obedience constrains the optimal solution through (\ref{eq:constraint_BOB_pi_V}).

If we suppress the constraint (\ref{eq:constraint_BOB_pi_V}), then the reduced optimization problem (\ref{eq:nonlinear_program_pi_object_Bayesian})-(\ref{eq:nonlinear_program_pi_constraint}) can be interpreted as a process to find pairs of decision variables $<\pi_{i}, V_{i}>$ that fit a Bellman optimality operator (i.e., satisfying (\ref{eq:Bellma_optimality_V})). 
In other words, the goal of this reduced optimization problem is to find fixed points.
However, there is another fixed point from the Bellman optimality operator established by the condition (\ref{eq:Bellma_optimality_J_Bayesian}) and the Bellman recursions (\ref{eq:bellman_V_1})-(\ref{eq:bellman_Q_1}); i.e., suppose agents are Bayesian obedient, for all $g\in\mathcal{G}$,$i\in\mathcal{N}$,
\begin{equation}\label{eq:bellman_optimality_J_not_V}
    \begin{aligned}
    &J_{i}(g|\bm{\theta}) = \max\limits_{\omega'_{i}} \sum_{ \bm{\omega}^{k}_{-i}, a_{i}}\bm{\alpha}_{-i}(\bm{\omega}^{k}_{-i}|g, \bm{\theta})\pi_{i}(a_{i}|g, \omega^{k}_{i}, \theta_{i})\\ &\times Q^{\bm{\pi}_{-i}}_{i}(a_{i}, g;\bm{\omega}^{k}|\bm{\theta};J_{i}).
    \end{aligned}
\end{equation}
However, the Bellman optimality equation (\ref{eq:bellman_optimality_J_not_V}) is independent of $V_{i}$ but is constructed based on the relationship between $J_{i}$ and $V_{i}$ given in (\ref{eq:bellman_J_1}).

We propose a design regime for finding a signaling rule $\bm{\alpha}$ that aligns two fixed points $J^{*}_{i}$ and $V^{*}_{i}$ for each agent $i$ while each agent's policy is strongly admissible.
Let, for any $g\in\mathcal{G}$, $\omega^{k}_{i}\in\Omega$ with $\alpha_{i}(\omega^{k}_{i}|g, \bm{\theta})>0$, $\omega_{i}\in\Omega$, $i\in\mathcal{N}$,
\begin{equation*}
    \begin{aligned}
    &V^{\bm{\alpha}_{-i}}_{i}(g,\omega_{i};\omega^{k}_{i}|\bm{\theta};V_{i})\\
    &\equiv \sum\limits_{\bm{\omega}^{k}_{-i}}\bm{\alpha}_{-i}(\bm{\omega}^{k}_{-i}|g,\bm{\theta}) V_{i}(g, \omega_{i}, \bm{\omega}^{k}_{-i}; \omega^{k}_{i}, \bm{\omega}^{k}_{-i}|\bm{\theta}),
    \end{aligned}
\end{equation*}
with $V^{\bm{\alpha}_{-i}}_{i}(g;\omega^{k}_{i}|\bm{\theta};V_{i}) = V^{\bm{\alpha}_{-i}}_{i}(g,\omega^{k}_{i};\omega^{k}_{i}|\bm{\theta};V_{i})$.
The objective function would be
\begin{equation}\label{eq:objective_function_FPA}
    \begin{aligned}
    &\bm{Z}^{\text{FPA}}(\bm{\alpha}, \bm{J}, \bm{V}; \bm{\theta}) \\
    &\equiv \sum\limits_{i,g}\Big( J_{i}(g|\bm{\theta})- \sum\limits_{\omega^{k}_{i}}\alpha_{i}(\omega^{k}_{i}|g, \bm{\theta}) V^{\bm{\alpha}_{-i}}_{i}(g;\omega^{k}_{i}|\bm{\theta})\Big),
    \end{aligned}
\end{equation}
which will need to be minimized by all possible $\bm{\alpha}$, $\bm{J}$, and $\bm{V}$.
%
%
By $\text{AD}[\bm{\alpha}, \bm{\kappa}]$, given $\bm{\alpha}$ and $\bm{\kappa}$, we define a set of valid policy profiles that are admissible. For example, if we refer to strong admissibility, the set $\text{AD}[\bm{\alpha}, \bm{\kappa}]$ is given as
\begin{equation}
    \begin{aligned}
    \text{AD}[\bm{\alpha}, \bm{\kappa}]\equiv&\Big\{\bm{\pi}: \forall i\in\mathcal{N}, \text{(\ref{eq:regular_policy})},\\ 
    &\bm{\kappa}(\bm{a}|g, \bm{\theta})\sum\limits_{\bm{\tilde{\omega}}_{-i} }\bm{\alpha}(\omega_{i}, \bm{\tilde{\omega}}_{-i}|g,\bm{\theta})\\
    &= \sum\limits_{\bm{\tilde{\omega}}_{-i} } \bm{\pi}(\bm{a}|g, \omega_{i}, \bm{\tilde{\omega}}_{-i} ,\bm{\theta})\bm{\alpha}(\omega_{i}, \bm{\tilde{\omega}}_{-i})\Big\}.
    \end{aligned}
\end{equation}
%
%
The Bayesian obedience is constrained by, for all $g\in\mathcal{G}$, $\omega_{i}\in\Omega$, $\omega^{k}_{i}\in\Omega$ with $\alpha_{i}(\omega^{k}_{i}|g, \bm{\theta})>0$, $i\in\mathcal{N}$,
\begin{equation}\tag{$\mathtt{BOB1}_{i}$}\label{eq:constraint_BOB_1}
    \begin{aligned}
    V^{\bm{\alpha}_{-i}}_{i}(g,\omega_{i};\omega^{k}_{i}|\bm{\theta};V_{i})\leq \sum\limits_{\omega^{k}_{i}} \bm{\alpha}({\omega}^{k}_{i}|g, \bm{\theta})V^{\bm{\alpha}_{-i}}_{i}(g;\omega^{k}_{i}|\bm{\theta};V_{i}).
    \end{aligned}
\end{equation}
Unlike the
which is conditioned on $V_{i}$ and $\bm{\alpha}$.
Unlike
The feasibility of $V_{i}$ given a $\bm{\pi}$ is captured by the constraint (\ref{eq:nonlinear_program_pi_constraint}). We additionally constrain the feasibility of $J_{i}$ in terms of $V_{i}$ as follows: for all $g\in\mathcal{G}$,  $\omega^{k}_{i}\in\Omega$ with $\alpha_{i}(\omega^{k}_{i}|g, \bm{\theta})>0$, $i\in\mathcal{N}$,
%
\begin{equation}\tag{$\mathtt{FS}_{i}$}\label{eq:FS_constraint_J}
    \begin{aligned}
    J_{i}(g|\bm{\theta})\geq \sum\limits_{\bm{\omega}^{k}_{-i}} \bm{\alpha}_{-i}(\bm{\omega}^{k}_{-i}|g, \bm{\theta})V_{i}(g, \omega^{k}_{i},\bm{\omega}^{k}_{-i}|\bm{\theta}).
    \end{aligned}
\end{equation}
Formally, the optimization problem of the information design based ob fixed-point alignment is
\begin{equation}\tag{$\mathtt{FPAlign}$}\label{eq:FPAlign}
    \begin{aligned}
    &\min \limits_{\bm{\alpha}, \bm{J}, \bm{V} } \bm{Z}^{\text{FPA}}(\bm{\alpha}, \bm{J}, \bm{V};\bm{\theta})\\
    \text{ s.t., }& \text{ (\ref{eq:nonlinear_program_pi_constraint}), (\ref{eq:constraint_BOB_1}), (\ref{eq:FS_constraint_J})}, \forall i\in\mathcal{N},\\
    &\bm{\pi}\in \text{AD}[\bm{\alpha}, \bm{\kappa}],  \bm{Z}(\bm{\pi}, \bm{V}; \bm{\alpha}) = 0. 
    \end{aligned}
\end{equation}
In (\ref{eq:FPAlign}), the constraint $Z(\bm{\pi}, \bm{V}; \bm{\alpha}) = 0$ is a sufficient and necessary for the feasible $\bm{\pi}$ to be the policy component of a Bayesian O-SMPE.

\begin{theorem}\label{thm:design_regime_first_version}
Fix a goal $\bm{\kappa}$. Let $<\bm{\alpha}^{*}, \bm{J}^{*}, \bm{V}^{*}>$ be feasible of (\ref{eq:FPAlign}).
The signaling rule $\bm{\alpha}^{*}$ is Bayesian OIL with strong admissibility if and only if \textit{(i)} $\bm{Z}^{\text{FPA}}(\bm{\alpha}^{*}, \bm{J}^{*}, \bm{V}^{*}; \bm{\theta})=0$, \textit{(ii)} $\bm{\kappa}\in \mathtt{MPG}[\bm{R}]$, and \textit{(iii)} strong admissibility holds.
\end{theorem}

Theorem \ref{thm:design_regime_first_version} provides a design regime for the signaling rule that is OIL in Bayesian O-SMPE. The condition \textit{(i)} specifies the optimality of the solution to (\ref{eq:FPAlign}) while the conditions \textit{(ii)} and \textit{(iii)} disciplines the principal's freedom in manipulating the agents' behaviors.
Specifically, the condition \textit{(ii)} implies that the principal cannot plan arbitrary goal that specifies arbitrary distribution of the agents' actions conditioning on the state and the joint type. 
%

Theorem \ref{thm:design_regime_first_version} shows two restrictions for the principal's freedom to set her goal and determine how the goal is achieved such that the agents' behaviors in a Bayesian O-SMPE can be influenced
Specifically, the goal $\bm{\kappa}$ should be a Markov perfect goal and the induced equilibrium policy profile should be strongly admissible.
The following corollary uncovers another restriction on the principal's ability to influence the agents' behaviors in a Bayesian O-SMPE.

\begin{corollary}\label{corollary:loss_of_generality_OIL}
Fix a base augmented Markov game $M$ with $\bm{R}=\{R_{i}\}_{i\in\mathcal{N}}$.
In general, there exists  $\bm{\kappa'}\in \mathtt{MPG}[\bm{R}]$ that can be achieved in an indirect game $M^{-D}[\bm{\alpha'}|\bm{\theta}]$ but not in any direct game $M^{D}[\bm{\alpha''}|\bm{\theta}]$.
\end{corollary}

Corollary \ref{corollary:loss_of_generality_OIL} states that restricting attention to direct information design is with loss of generality in selecting Markov perfect goals.

\section{Principal's Optimal Information Design}

So far, we focus on when the principal's goal is given.
In this section, we introduce the optimality criterion of the principal's goal selection and define the optimal information design problem without a predetermined goal.
The one-stage payoff function of the principal is $u(\cdot;\bm{\theta}): \mathcal{A}^{n}\times \mathcal{G} \mapsto \mathbb{R}$, such that $u(\bm{a}, g; \bm{\theta})$ gives the immediate payoff for the principal when the state is $g$ and the agents of the game $M[\bm{\alpha};\bm{\theta}]$ take the joint action $\bm{a}$.
Recall that the principal's goal $\bm{\kappa}$ is the probability distribution of the agents' joint action in the equilibrium conditioned only on the global state given the agents' types.
Hence, the information structure that matters for the principal's goal selection problem is $<\mathcal{G}, \mathcal{T}_{g}, d_{g}>$.
According to Ionescu Tulcea theorem, the information structure $<\mathcal{G}, \mathcal{T}_{g}, d_{g}>$ and any goal $\bm{\kappa}$ uniquely define a probability measure on $(\mathcal{G} \times \mathcal{A}^{n})^{\infty}$.
We denote the corresponding expectation as $\mathbb{E}^{\bm{\kappa}}[\cdot]$. 
Hence, the principal's problem is to choose a goal by maximizing her expected payoff ($\gamma$-discounted, the same as the agents'), i.e.,
\begin{equation}\label{eq:principal_optimal_info_design_v1}
    \begin{aligned}
     C(\bm{\kappa})\equiv \mathbb{E}^{\bm{\kappa}}\Big[\sum_{t=0}^{\infty}\gamma^{t} u(\bm{a}_{t}, g_{t}, \bm{\theta})\Big].
    \end{aligned}
\end{equation}
However, the principal cannot force the agents to take the actions or directly program agents' actions according to the $\bm{\kappa}^{*}$ that maximizes $C(\bm{\kappa})$; instead, she uses information design to elicit the agents to take actions that coincide with $\bm{\kappa}^{*}$ in the sense of strong admissibility.
Hence, the principal's optimal goal selection problem is a constrained optimization problem:
\begin{equation}\label{eq:principal_goal_problem_1}
    \begin{aligned}
    &\max\limits_{\bm{\kappa}\in\mathtt{MPG}[\bm{R}]} C(\bm{\kappa})\equiv \mathbb{E}^{\bm{\kappa}}\Big[\sum_{t=0}^{\infty}\gamma^{t} u(\bm{a}_{t}, g_{t}, \bm{\theta})\Big]\\
    \text{s.t. }& <\bm{\alpha}^{*}, \bm{J}^{*}, \bm{V}^{*}> \text{ is a solution of (\ref{eq:FPAlign})}.
    \end{aligned}
\end{equation}

In (\ref{eq:principal_goal_problem_1}), the feasibility of $\bm{\kappa}$ is captured by the conditions \textit{(i)} it is a Markov perfect goal and \textit{(ii)} it disciplines the strong admissibility in (\ref{eq:FPAlign}).

The optimal goal selection problem (\ref{eq:principal_goal_problem_1}) can be reformulated to a problem of selecting $\bm{\alpha}$ and $\bm{\pi}$.
Specifically, from the (strong) admissibility, the objective function $C(\bm{\kappa})$ can be represented in terms of $\bm{\alpha}$ and $\bm{\pi}$ as follows:
\begin{equation}\label{eq:transformed_principal}
    \begin{aligned}
    &C^{O}(\bm{\alpha}, \bm{\pi})\equiv \mathbb{E}\Big[\sum^{\infty}_{t=0}\sum\limits_{\bm{a}_{t}}\\
    &\gamma^{t}u(\bm{a}_{t}, g_{t};\bm{\theta})\sum_{\bm{\omega}^{k}_{t} }\bm{\pi}(\bm{a}_{t}|g_{t}, \bm{\omega}^{k}_{t}, \bm{\theta})\bm{\alpha}(\bm{\omega}^{k}_{t}|g_{t},\bm{\theta})\Big],
    \end{aligned}
\end{equation}
where the expectation $\mathbb{E}$ is with respect to the probability measure on the dynamics of the state.
%
%
Let $\bm{\Pi}[\bm{\alpha}, \bm{J},\bm{V}]$ denote the set of valid policy profiles that associated with the value function $\bm{V}$, given $\bm{\alpha}$ and $\bm{J}$; i.e., 
\begin{equation*}
    \begin{aligned}
    \bm{\Pi}[\bm{\alpha},& \bm{J},\bm{V}]\equiv\Big\{\bm{\pi}: \text{(\ref{eq:regular_policy})}, V_{i}(g; \bm{\omega}^{k}|\bm{\theta})\\
    &= \mathbb{E}_{\pi_{i}}\Big[ Q^{\bm{\pi}_{-i}}_{i}(a_{i}, g, \omega_{i}, \bm{\omega}^{k}_{-i};\bm{\omega}^{k}|\bm{\theta};J_{i}) \Big],  \forall i\in\mathcal{N}\Big\},
    \end{aligned}
\end{equation*}
where $Q^{\bm{\pi}_{-i}}_{i}$ is defined in (\ref{eq:pi_specified_Q}).
Hence, the principal's problem (\ref{eq:principal_goal_problem_1}) can be reformulated as follows:
\begin{equation}\tag{OptInfo}\label{eq:principal_optimal_inforD}
    \begin{aligned}
    &\max\limits_{\bm{\alpha}} \max\limits_{\bm{\pi}\in \bm{\Pi}[\bm{\alpha}^{*}, \bm{J}^{*},\bm{V}^{*}] } C^{O}(\bm{\alpha}, \bm{\pi})\\
    \text{s.t. }& 
    <\bm{\alpha}, \bm{J}^{*}, \bm{V}^{*}> \in\argmin \limits_{\bm{\alpha}, \bm{J}, \bm{V} } \bm{Z}^{\text{FPA}}(\bm{\alpha}, \bm{J}, \bm{V};\bm{\theta})\\
    &\text{s.t. } \text{ (\ref{eq:nonlinear_program_pi_constraint}), (\ref{eq:constraint_BOB_1}), (\ref{eq:FS_constraint_J})}, \forall i\in\mathcal{N}.
    \end{aligned}
\end{equation}
Technically, the problem (\ref{eq:principal_optimal_inforD}) is to select \textit{(i)} an equilibrium policy profile $\bm{\pi}^{*}$ that is strongly admissible and is a MPE and \textit{(ii)} the signaling rule $\bm{\alpha}^{*}$ that induces the policy profile such that the principal's expected payoff $C^{O}$ is maximized at $(\bm{\alpha}^{*}, \bm{\pi}^{*})$.

The problem (\ref{eq:principal_optimal_inforD}) is, however, based on the assumption that the agents' equilibrium behavior is always principal-preferred.  
We could also consider the problem of a principal who aims to solve her problem in a robust manner in the sense that she chooses the signaling rule, but wants to maximize her expected payoff in the worst equilibrium; i.e., it is the robust information design problem:
\begin{equation}\tag{Robust}\label{eq:principal_robust_inforD}
    \begin{aligned}
    &\max\limits_{\bm{\alpha}} \min\limits_{\bm{\pi}\in \bm{\Pi}[\bm{\alpha}^{*}, \bm{J}^{*},\bm{V}^{*}] } C^{O}(\bm{\alpha}, \bm{\pi})\\
    \text{s.t. }& 
    <\bm{\alpha}, \bm{J}^{*}, \bm{V}^{*}> \in\argmin \limits_{\bm{\alpha}, \bm{J}, \bm{V} } \bm{Z}^{\text{FPA}}(\bm{\alpha}, \bm{J}, \bm{V};\bm{\theta})\\
    &\text{s.t. } \text{ (\ref{eq:nonlinear_program_pi_constraint}), (\ref{eq:constraint_BOB_1}), (\ref{eq:FS_constraint_J})}, \forall i\in\mathcal{N}.
    \end{aligned}
\end{equation}

\subsection{Fixed-Point Misalignment Minimization}


%

In this section, we provide an alternative formulation of information design by introducing the notion of \textit{fixed-point misalignment} (FP misalignment).

Define, for any $g\in\mathcal{G}$, $i\in\mathcal{N}$, $\bm{\alpha}_{-i}$, $\bm{\pi}_{-i}$, $J_{i}$, $V_{i}$,
\begin{equation*}
    \mathcal{E}^{\bm{\alpha}_{-i} }_{i}(J_{i}, V_{i}; g, \omega^{k}_{i},\bm{\theta} ) \equiv J_{i}(g|\bm{\theta}) - V^{\bm{\alpha}_{-i}}_{i}(g;\omega^{k}_{i}|\bm{\theta}; V_{i}),
\end{equation*}
\begin{equation*}
    \mathcal{E}^{\bm{\pi}_{-i} }_{i}(J_{i}, V_{i}; g, \bm{\omega}^{k},\bm{\theta} ) \equiv V_{i}(g;\bm{\omega}^{k}|\bm{\theta}) - Q^{\bm{\pi}_{-i}}_{i}(a_{i}, g; \bm{\omega}^{k}|\bm{\theta}; J_{i}).
\end{equation*}
Then, we define the notion of \textit{fixed-point misalignment} as follows:
\begin{equation*}
    \begin{aligned}
    \delta^{\bm{\pi}_{-i}}_{i}(&J_{i}, V_{i};\pi_{i}|g, \omega^{k}_{i},a_{i},\bm{\theta})\\
    &\equiv \pi_{i}(a_{i}|g,\omega^{k}_{i}, \theta_{i})\mathcal{E}^{\bm{\pi}_{-i} }_{i}(J_{i}, V_{i}; g, \bm{\theta} ),
    \end{aligned}
\end{equation*}
\begin{equation*}
    \begin{aligned}
    \delta^{\bm{\alpha}_{-i}}_{i}(& J_{i}, V_{i}; \alpha_{i}|g, \omega^{k}_{i},\bm{\theta})\\
    &\equiv \alpha_{i}(\omega^{k}_{i}|g,\omega^{k}_{i},\bm{\theta})\mathcal{E}^{\bm{\alpha}_{-i} }_{i}(J_{i}, V_{i}; g,\bm{\omega}^{k}, \bm{\theta}).
    \end{aligned}
\end{equation*}

\begin{proposition}\label{prop:design_regime_second_version}
Fix a Markov perfect goal $\bm{\kappa}$.
A strategy profile $<\bm{\beta}^{O}, \bm{\pi}^{*}>$ where $\bm{\beta}^{O}$ is Bayesian obedient is a Bayesian O-SMPE if and only if there exists a profile $<\bm{\alpha}, \bm{J}, \bm{V}>$ that satisfies (\ref{eq:nonlinear_program_pi_constraint}), (\ref{eq:constraint_BOB_1}), and (\ref{eq:FS_constraint_J}), given $\bm{\pi}^{*}$, $g\in\mathcal{G}$,  $i\in\mathcal{N}$, such that, for all $\omega^{k}_{i}\in\Omega$, $a_{i}\in\mathcal{A}$, $i\in\mathcal{N}$,
\begin{equation}\tag{$\mathtt{FPM1}_{i}$}\label{eq:misalignment_1}
    \begin{aligned}
    \delta^{\bm{\pi}_{-i}}_{i}(J_{i}, V_{i};\pi_{i}|g, \omega^{k}_{i},a_{i},\bm{\theta})=0,
    \end{aligned}
\end{equation}
\begin{equation}\tag{$\mathtt{FPM2}_{i}$}\label{eq:misalignment_2}
    \begin{aligned}
    \delta^{\bm{\alpha}_{-i}}_{i}( J_{i}, V_{i};\alpha_{i}|g, \omega^{k}_{i},\bm{\theta})=0.
\end{aligned}
\end{equation}
\end{proposition}

Then, we reformulate (\ref{eq:FPAlign}) in terms of FP misalignment minimization based on Proposition \ref{prop:design_regime_second_version}.
Due to the definitions of $\mathcal{E}^{\bm{\alpha}_{-i}}_{i}$ and $\mathcal{E}^{\bm{\pi}_{-i}}_{i}$, the objective functions $\bm{Z}$ and  $\bm{Z}^{FPA}$ can be represented in terms of $\delta^{\bm{\pi}}_{i}$ and $\delta^{\bm{\alpha}}_{i}$, as follows (denoted by $\bm{\hat{Z}}$ and $\bm{\hat{Z}}^{FPA}$) respectively:
\begin{equation}
    \begin{aligned}
    \bm{\hat{Z}}(\bm{\pi}, \bm{V}; \bm{\alpha})=\sum\limits_{i,g, \omega^{k}_{i}}\sum\limits_{a_{i}}\delta^{\bm{\pi}_{-i}}_{i}(J_{i}, V_{i}; \pi_{i}|g, \omega^{k}_{i},a_{i},\bm{\theta}),
    \end{aligned}
\end{equation}
\begin{equation}
    \begin{aligned}
    \bm{\hat{Z}}^{FPA}(\bm{\alpha}, \bm{J}, \bm{V}; \bm{\theta}) = \sum\limits_{i,g}\sum\limits_{\omega^{k}_{i}}\delta^{\bm{\alpha}_{-i}}_{i}( J_{i}, V_{i};\alpha_{i}|g, \omega^{k}_{i},\bm{\theta}).
    \end{aligned}
\end{equation}

\begin{corollary}
Given any $\bm{\kappa}\in\mathtt{MPG}[\bm{R}]$, the problem (\ref{eq:FPAlign}) is equivalent to the following:
\begin{equation}\tag{$\mathtt{FPMis}$}\label{eq:FPMis}
    \begin{aligned}
    \min\limits_{\bm{\alpha}, \bm{J},\bm{V}}&\bm{\hat{Z}}^{FPA}(\bm{\alpha}, \bm{J}, \bm{V}; \bm{\theta}) \\
    \text{s.t. }& \text{ (\ref{eq:constraint_BOB_1}) ,(\ref{eq:misalignment_1}) ,(\ref{eq:misalignment_2}), } \forall i\in\mathcal{N}\\
    & \bm{\pi}\in \text{AD}[\bm{\alpha},\bm{\kappa}].
    \end{aligned}
\end{equation}
\end{corollary}

Define a set:
\begin{equation*}
    \begin{aligned}
    \bm{\mathcal{D}}\equiv\Big\{ \bm{\alpha}, \bm{\pi}: \forall i\in\mathcal{N},& \text{(\ref{eq:regular_policy})},\\
     <\alpha_{i}, J_{i}, V_{i}>\in&\argmin \delta^{\bm{\alpha}_{-i}}_{i}( J_{i}, V_{i};\alpha_{i}|g, \omega^{k}_{i},\bm{\theta})\\
    &\text{s.t. } \text{ (\ref{eq:constraint_BOB_1}) ,(\ref{eq:misalignment_1}) ,(\ref{eq:misalignment_2})}.\Big\}
    \end{aligned}
\end{equation*}
%
%
%
Given $\bm{\mathcal{D}}$, we define a set of OIL signaling rules as $\bm{\mathcal{S}}\equiv \Big\{\bm{\alpha}: \forall (\bm{\alpha}, \bm{\pi}) \in \bm{\mathcal{D}}\Big\}$
and a set of policy profiles given any signaling rule $\bm{\alpha}$ as $\bm{\Pi}[\bm{\alpha}]\equiv\Big\{\bm{\pi}:  \forall (\bm{\alpha}, \bm{\pi}) \in \bm{\mathcal{D}} \Big\}$.

\begin{corollary}
The principal's robust information design is to solve the following problem
\begin{equation}\label{eq:robust_info_D_MFPM}
    \begin{aligned}
    \max\limits_{\bm{\alpha}\in \bm{\mathcal{S}}} \min\limits_{\bm{\pi}\in \bm{\Pi}[\bm{\alpha}]} C^{O}(&\bm{\alpha}, \bm{\pi}).
    \end{aligned}
\end{equation}
\end{corollary}


\section{Conclusion}

This work is the first to propose an information design principle for dynamic games in which each agent makes coupled decisions of selecting a signal and taking an action at each period of time.
We have formally defined a novel information design problem for the indirect and the direct settings.
The notion of obedient implementability has been introduced to capture the optimality of the direct information design problem in a new equilibrium concept of obedient sequential Markov perfect equilibrium (O-SMPE).
By characterizing the obedient implementability (OIL) in Bayesian O-SMPE, we have proposed an approach to determining the information structure. 
We refer to this approach as fixed-point alignment that aligns the two fixed points at the signal selection stage and the action taken stage, respectively.
We have uncovered the restrictions that discipline the principal's freedom to influence the agents' behaviors in Bayesian O-SMPE.
Specifically, the principal's goal should be a Markov perfect goal and the equilibrium policy profile should be strongly admissible. Additionally, it is with loss of generality in terms of the selection of Markov perfect goals for OIL in Bayesian O-SMPE.
Finally, we have formulated the principal's goal selection problem in terms of the optimal and the robust information design by replacing the admissibility by the optimality or the robustness of the agents' equilibrium policy profile in the principal's expected payoff.



\bibliographystyle{IEEEtran}
\bibliography{infoDMAS}

\begin{thebibliography}{10}
\providecommand{\url}[1]{#1}
\csname url@samestyle\endcsname
\providecommand{\newblock}{\relax}
\providecommand{\bibinfo}[2]{#2}
\providecommand{\BIBentrySTDinterwordspacing}{\spaceskip=0pt\relax}
\providecommand{\BIBentryALTinterwordstretchfactor}{4}
\providecommand{\BIBentryALTinterwordspacing}{\spaceskip=\fontdimen2\font plus
\BIBentryALTinterwordstretchfactor\fontdimen3\font minus
  \fontdimen4\font\relax}
\providecommand{\BIBforeignlanguage}[2]{{%
\expandafter\ifx\csname l@#1\endcsname\relax
\typeout{** WARNING: IEEEtran.bst: No hyphenation pattern has been}%
\typeout{** loaded for the language `#1'. Using the pattern for}%
\typeout{** the default language instead.}%
\else
\language=\csname l@#1\endcsname
\fi
#2}}
\providecommand{\BIBdecl}{\relax}
\BIBdecl

\bibitem{dickinson1985actions}
A.~Dickinson, ``Actions and habits: the development of behavioural autonomy,''
  \emph{Philosophical Transactions of the Royal Society of London. B,
  Biological Sciences}, vol. 308, no. 1135, pp. 67--78, 1985.

\bibitem{bergemann2019information}
D.~{Bergemann} and S.~{Morris}, ``Information design: A unified perspective,''
  \emph{Journal of Economic Literature}, vol.~57, no.~1, pp. 44--95, 2019.

\bibitem{taneva2019information}
I.~Taneva, ``Information design,'' \emph{American Economic Journal:
  Microeconomics}, vol.~11, no.~4, pp. 151--85, 2019.

\bibitem{chentanez2005intrinsically}
N.~Chentanez, A.~G. Barto, and S.~P. Singh, ``Intrinsically motivated
  reinforcement learning,'' in \emph{Advances in neural information processing
  systems}, 2005, pp. 1281--1288.

\bibitem{mathevet2020information}
L.~Mathevet, J.~Perego, and I.~Taneva, ``On information design in games,''
  \emph{Journal of Political Economy}, vol. 128, no.~4, pp. 1370--1404, 2020.

\bibitem{bergemann2016bayes}
D.~Bergemann and S.~Morris, ``Bayes correlated equilibrium and the comparison
  of information structures in games,'' \emph{Theoretical Economics}, vol.~11,
  no.~2, pp. 487--522, 2016.

\bibitem{kamenica2011bayesian}
E.~Kamenica and M.~Gentzkow, ``Bayesian persuasion,'' \emph{American Economic
  Review}, vol. 101, no.~6, pp. 2590--2615, 2011.

\bibitem{ely2015suspense}
J.~Ely, A.~Frankel, and E.~Kamenica, ``Suspense and surprise,'' \emph{Journal
  of Political Economy}, vol. 123, no.~1, pp. 215--260, 2015.

\bibitem{passadore2015robust}
J.~Passadore and J.~P. Xandri, ``Robust conditional predictions in dynamic
  games: An application to sovereign debt,'' \emph{Job Market Paper}, 2015.

\bibitem{doval2020sequential}
L.~Doval and J.~C. Ely, ``Sequential information design,'' \emph{Econometrica},
  vol.~88, no.~6, pp. 2575--2608, 2020.

\bibitem{ely2017beeps}
J.~C. Ely, ``Beeps,'' \emph{American Economic Review}, vol. 107, no.~1, pp.
  31--53, 2017.

\bibitem{ely2020moving}
J.~C. Ely and M.~Szydlowski, ``Moving the goalposts,'' \emph{Journal of
  Political Economy}, vol. 128, no.~2, pp. 468--506, 2020.

\bibitem{makris2018information}
M.~Makris and L.~Renou, ``Information design in multi-stage games,'' working
  paper, Tech. Rep., 2018.

\bibitem{koessler2018interactive}
F.~Koessler, M.~Laclau, and T.~Tomala, ``Interactive information design,''
  \emph{HEC Paris Research Paper No. ECO/SCD-2018-1260}, 2018.

\bibitem{myerson1981optimal}
R.~B. Myerson, ``Optimal auction design,'' \emph{Mathematics of operations
  research}, vol.~6, no.~1, pp. 58--73, 1981.

\bibitem{pavan2014dynamic}
A.~Pavan, I.~Segal, and J.~Toikka, ``Dynamic mechanism design: A myersonian
  approach,'' \emph{Econometrica}, vol.~82, no.~2, pp. 601--653, 2014.

\bibitem{zhang2021differential}
T.~Zhang and Q.~Zhu, ``On the differential private data market: Endogenous
  evolution, dynamic pricing, and incentive compatibility,'' 2021.

\bibitem{milgrom2004putting}
P.~Milgrom and P.~R. Milgrom, \emph{Putting auction theory to work}.\hskip 1em
  plus 0.5em minus 0.4em\relax Cambridge University Press, 2004.

\bibitem{bhat2019optimal}
S.~Bhat, S.~Jain, S.~Gujar, and Y.~Narahari, ``An optimal bidimensional
  multi-armed bandit auction for multi-unit procurement,'' \emph{Annals of
  Mathematics and Artificial Intelligence}, vol.~85, no.~1, pp. 1--19, 2019.

\bibitem{sonmez2011matching}
T.~S{\"o}nmez and M.~U. {\"U}nver, ``Matching, allocation, and exchange of
  discrete resources,'' in \emph{Handbook of social Economics}.\hskip 1em plus
  0.5em minus 0.4em\relax Elsevier, 2011, vol.~1, pp. 781--852.

\bibitem{zhang2019optimal}
T.~Zhang and Q.~Zhu, ``Optimal two-sided market mechanism design for
  large-scale data sharing and trading in massive iot networks,'' \emph{arXiv
  preprint arXiv:1912.06229}, 2019.

\bibitem{dewey2014reinforcement}
D.~Dewey, ``Reinforcement learning and the reward engineering principle,'' in
  \emph{2014 AAAI Spring Symposium Series}, 2014.

\bibitem{nagpal2020reward}
R.~Nagpal, A.~U. Krishnan, and H.~Yu, ``Reward engineering for object pick and
  place training,'' \emph{arXiv preprint arXiv:2001.03792}, 2020.

\bibitem{hadfield2017inverse}
D.~Hadfield-Menell, S.~Milli, P.~Abbeel, S.~J. Russell, and A.~Dragan,
  ``Inverse reward design,'' in \emph{Advances in neural information processing
  systems}, 2017, pp. 6765--6774.

\bibitem{kamenica2019bayesian}
E.~Kamenica, ``Bayesian persuasion and information design,'' \emph{Annual
  Review of Economics}, vol.~11, pp. 249--272, 2019.

\bibitem{brocas2007influence}
I.~Brocas and J.~D. Carrillo, ``Influence through ignorance,'' \emph{The RAND
  Journal of Economics}, vol.~38, no.~4, pp. 931--947, 2007.

\bibitem{rayo2010optimal}
L.~Rayo and I.~Segal, ``Optimal information disclosure,'' \emph{Journal of
  political Economy}, vol. 118, no.~5, pp. 949--987, 2010.

\bibitem{arieli2019private}
I.~Arieli and Y.~Babichenko, ``Private bayesian persuasion,'' \emph{Journal of
  Economic Theory}, vol. 182, pp. 185--217, 2019.

\bibitem{castiglioni2020online}
M.~Castiglioni, A.~Celli, A.~Marchesi, and N.~Gatti, ``Online bayesian
  persuasion,'' \emph{Advances in Neural Information Processing Systems},
  vol.~33, 2020.

\bibitem{mertens1985formulation}
J.-F. Mertens and S.~Zamir, ``Formulation of bayesian analysis for games with
  incomplete information,'' \emph{International Journal of Game Theory},
  vol.~14, no.~1, pp. 1--29, 1985.

\bibitem{goldstein2018stress}
I.~Goldstein and Y.~Leitner, ``Stress tests and information disclosure,''
  \emph{Journal of Economic Theory}, vol. 177, pp. 34--69, 2018.

\bibitem{inostroza2018persuasion}
N.~Inostroza and A.~Pavan, ``Persuasion in global games with application to
  stress testing,'' 2018.

\bibitem{hernandez2018bayesian}
P.~Hern{\'a}ndez and Z.~Neeman, ``How bayesian persuasion can help reduce
  illegal parking and other socially undesirable behavior,'' \emph{Preprint},
  2018.

\bibitem{rabinovich2015information}
Z.~Rabinovich, A.~X. Jiang, M.~Jain, and H.~Xu, ``Information disclosure as a
  means to security,'' in \emph{Proceedings of the 2015 International
  Conference on Autonomous Agents and Multiagent Systems}.\hskip 1em plus 0.5em
  minus 0.4em\relax Citeseer, 2015, pp. 645--653.

\bibitem{gehlbach2014government}
S.~Gehlbach and K.~Sonin, ``Government control of the media,'' \emph{Journal of
  public Economics}, vol. 118, pp. 163--171, 2014.

\bibitem{das2017reducing}
S.~Das, E.~Kamenica, and R.~Mirka, ``Reducing congestion through information
  design,'' in \emph{2017 55th annual allerton conference on communication,
  control, and computing (allerton)}.\hskip 1em plus 0.5em minus 0.4em\relax
  IEEE, 2017, pp. 1279--1284.

\bibitem{duffie2017benchmarks}
D.~Duffie, P.~Dworczak, and H.~Zhu, ``Benchmarks in search markets,'' \emph{The
  Journal of Finance}, vol.~72, no.~5, pp. 1983--2044, 2017.

\bibitem{szydlowski2021optimal}
M.~Szydlowski, ``Optimal financing and disclosure,'' \emph{Management Science},
  vol.~67, no.~1, pp. 436--454, 2021.

\bibitem{garcia2021information}
D.~Garcia and M.~Tsur, ``Information design in competitive insurance markets,''
  \emph{Journal of Economic Theory}, vol. 191, p. 105160, 2021.

\bibitem{ziebart2011maximum}
B.~D. Ziebart, J.~A. Bagnell, and A.~K. Dey, ``Maximum causal entropy
  correlated equilibria for markov games.'' in \emph{AAMAS}.\hskip 1em plus
  0.5em minus 0.4em\relax Citeseer, 2011, pp. 207--214.

\bibitem{hernandez2012discrete}
O.~Hern{\'a}ndez-Lerma and J.~B. Lasserre, \emph{Discrete-time Markov control
  processes: basic optimality criteria}.\hskip 1em plus 0.5em minus 0.4em\relax
  Springer Science \& Business Media, 2012, vol.~30.

\bibitem{he2017stationary}
W.~He and Y.~Sun, ``Stationary markov perfect equilibria in discounted
  stochastic games,'' \emph{Journal of Economic Theory}, vol. 169, pp. 35--61,
  2017.

\bibitem{bellman1966dynamic}
R.~Bellman, ``Dynamic programming,'' \emph{Science}, vol. 153, no. 3731, pp.
  34--37, 1966.

\bibitem{filar1997competitive}
J.~Filar and K.~Vrieze, ``Competitive markov decision processes-theory,
  algorithms, and applications,'' 1997.

\bibitem{prasad2012general}
H.~Prasad and S.~Bhatnagar, ``General-sum stochastic games: Verifiability
  conditions for nash equilibria,'' \emph{Automatica}, vol.~48, no.~11, pp.
  2923--2930, 2012.

\bibitem{prasad2015two}
H.~Prasad, P.~LA, and S.~Bhatnagar, ``Two-timescale algorithms for learning
  nash equilibria in general-sum stochastic games,'' in \emph{Proceedings of
  the 2015 International Conference on Autonomous Agents and Multiagent
  Systems}, 2015, pp. 1371--1379.

\end{thebibliography}
%



\end{document}